\documentclass[sigconf,nonacm]{acmart}

\setcopyright{cc}
\copyrightyear{2026}
\acmYear{2026}
\acmDOI{XXXXXXX.XXXXXXX}
\acmConference[Conf]{Conference}{2026}{}

\usepackage{makecell}
\usepackage{multirow}
\usepackage{subcaption}
\usepackage{todonotes}
\usepackage[table]{xcolor}
\usepackage{colortbl}
\usepackage{array}

\makeatletter
\newcommand{\defnote}[2]{%
  \refstepcounter{footnote}%
  \label{#1}%
  \footnotetext{#2}%
}
\newcommand{\tabfootref}[1]{\textsuperscript{\hyperref[#1]{\ref*{#1}}}}
\makeatother

\definecolor{columngray}{gray}{0.9} %
\definecolor{lightgray}{gray}{0.9}
\definecolor{verylightgray}{gray}{0.95}
\newcolumntype{G}{>{\hspace{-\tabcolsep}\columncolor{lightgray}}c<{\hspace{-\tabcolsep}}}
\newcommand{\head}[1]{\par\noindent\textbf{#1:}\space}

\begin{document}

\title{The Wisdom of the Crowd: LLM-Ensemble Code Generation and Repair}

\title{Applying the Wisdom of the Crowd to LLM-based Code Generation and Repair}  %

\title{The Wisdom of the Crowd in LLM-based Code Generation and Repair}  %

\title[Wisdom and Delusion of LLM Ensembles for Code Generation and Repair]{Wisdom and Delusion of LLM Ensembles\\for Code Generation and Repair}

\newcommand{\SimulaAffiliation}{\affiliation{%
  \institution{Simula Research Laboratory}%
  \city{Oslo}%
  \country{Norway}%
}}

\author{Fernando Vallecillos-Ruiz}
\email{fernando@simula.no}
\orcid{0000-0001-7213-3732}
\SimulaAffiliation{}
\author{Max Hort}
\orcid{0000-0001-8684-5909}
\email{maxh@simula.no}
\SimulaAffiliation{}
\author{Leon Moonen}
\orcid{0000-0002-1761-6771}
\email{leon.moonen@computer.org}
\SimulaAffiliation{}

\begin{abstract}
Today's pursuit of a single Large Language Model (LMM) for all software engineering tasks is resource-intensive and overlooks the potential benefits of complementarity, where different models contribute unique strengths.
However, the degree to which coding LLMs complement each other and the best strategy for maximizing an ensemble's potential are unclear, leaving practitioners without a clear path to move beyond single-model systems.

To address this gap, we empirically compare ten individual LLMs from five families, and three ensembles of these LLMs 
across three software engineering benchmarks covering code generation and program repair.
We assess the complementarity between models and the performance gap between the best individual model and the ensembles.
Next, we evaluate various selection heuristics to identify correct solutions from an ensemble's candidate pool.

\begin{sloppypar}
We find that the theoretical upperbound for an ensemble's performance can be 83\% above the best single model. 
Our results show that consensus-based strategies for selecting solutions fall into a ``\emph{popularity trap},'' amplifying common but incorrect outputs.
In contrast, a diversity-based strategy realizes up to 95\% of this theoretical potential,
and proves effective even in small two-model ensembles, 
enabling a cost-efficient way to enhance performance by leveraging multiple LLMs.   
\end{sloppypar}
\end{abstract}


\maketitle
\section{Introduction}

Large Language Models (LLMs) have demonstrated remarkable skill in software engineering tasks, from code generation to Automatic Program Repair (APR)~\cite{fan2023:large, hou2024:large}.
Their ability to generate human-like code has promised to enhance developer productivity and accelerate software development~\cite{rasheed2025:large}.
This has resulted in an arms race to investigate and develop a single, superior model that can outperform all others across any benchmark~\cite{white2024:livebench, du2024:evaluating}, aiming to find a definite answer to a question that has become more and more common: 
``Which LLM is the best for coding-related tasks?''
The core problem with this approach is the implicit assumption that a single model can eventually dominate all others across the heterogeneous landscape of programming problems~\cite{dou2024:whats,shypula2025:evaluating}.

The search for a single best model overlooks experience in related fields,
such as in the pre-LLM era of APR. 
It
was well accepted 
that the landscape of bugs was too diverse for any single repair tool to tackle
~\cite{zhong2024:practical}.
State-of-the-art results were achieved by creating ensembles of diverse techniques or tools (e.g. combining search-based, constraint-based, and template-based approaches) where the collective result exceeded that of any individual component.
We argue that this 
principle of complementarity
is still 
fundamental
to modern 
LLMs,
where no single model can solve all problems.
Models trained on diverse datasets with varying architectures or numbers of parameters likely develop unique problem-solving capabilities.
However, the field currently lacks a clear understanding of the degree to which each model solves a unique set of problems that other models cannot.
Recently, 
the need for fundamental understanding has been 
rising
due to the 
increased interest in
multi-agent systems where multiple LLM-based agents collaborate on complex software engineering tasks~\cite{he2025:llmbased}.
The rationale behind multi-agent systems and ensembles of models is that the combination of multiple models can overcome their individual limitations and outperform any single model~\cite{mienye2022:survey}.

\begin{figure}
    \centering
    \includegraphics[width=\linewidth]{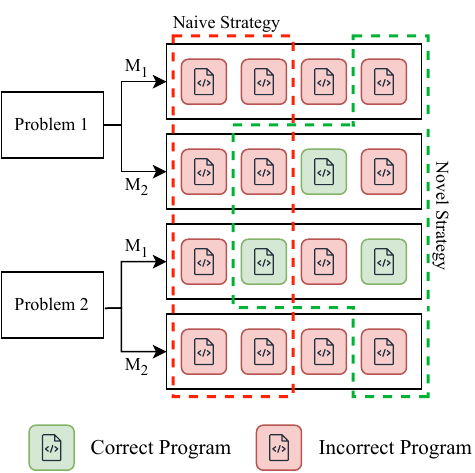}
    \caption{Conceptual overview of the selection process from an LLM ensemble's output pool where only 4 programs can be selected per problem.}
    \label{fig:overview-strategy-ensemble}
\end{figure}

Establishing model complementarity is necessary, but not 
the only 
requirement to obtain practical benefits.
Each model in an ensemble generates multiple candidates, creating a noisy pool of potential solutions.
Naive selection from this pool can be ineffective.
Intuitive heuristics, such as selecting the most common solution might seem promising, but could lead models to converge on plausible but incorrect answers.
Alternatively, one could leverage confidence scores of each model, but the effectiveness of these signals coming from models with different architectures remains an open question.
Without a systematic evaluation, the 
practical
benefits of model complementarity will remain theoretical, leaving practitioners without a reliable method to realize this potential. 
\autoref{fig:overview-strategy-ensemble} shows that while model complementarity can create a pool of candidates containing solutions for two problems, only an effective selection heuristic can realize that potential.

Therefore, we perform an empirical study to systematically explore and quantify the potential of LLM ensembles for software engineering tasks.
Our goal is not to compete for a new state-of-the-art score on a benchmark, but rather to provide fundamental understanding on ensembles and how they compare to a single-model paradigm.
To achieve this, we select a set of ten instruction-tuned LLMs with strong coding capabilities, spanning five different model families.
This set consists of five smaller models ($\sim$7B parameters) and five larger models ($\sim$14B parameters).
We evaluate these models across three common benchmarks covering two software engineering tasks: HumanEval-Java~\cite{jiang2023:impact} and Defects4J~\cite{just2014:defects4j} for APR, and LiveCodeBench~\cite{jain2024:livecodebench} covering code generation.

Our investigation consists of two phases.
First, we establish a theoretical foundation for ensembles by analyzing model complementarity among models in the same family 
and those with different parameter counts.
This allows us to quantify the performance gap between an individual model and the theoretical potential of an ensemble.
Secondly, having established this performance ceiling, we propose and evaluate a set of heuristics to select candidates from the aggregated output pool as illustrated in \autoref{fig:overview-strategy-ensemble}.
Through this analysis, we aim to provide concrete, empirically driven strategies to effectively realize the potential of multiple LLMs.

Our contributions are as follows:
\begin{itemize}
    \item We conduct an extensive empirical study with ten LLMs from five families across three code-related benchmarks, quantifying and analyzing for the first time the performance gap between single models and ensembles for software engineering tasks.
    \item We provide empirical evidence that LLMs exhibit significant complementarity 
    where even smaller models contribute unique solutions not found by large state-of-the-art models.
    \item We quantify the performance ceiling offered by ensembles of models, demonstrating a potential performance increase of up to 83\% in the number of problems solved by the best individual model.
    \item We systematically evaluate various selection heuristics to identify correct solutions from a pool of candidates.
    \item We identify the challenge of the ``\emph{popularity trap}'' where models frequently converge on the same incorrect solution and provide a strategy to counter this trap that realizes up to 95\% of the ensemble's theoretical potential.
    \item After initial experiments with ensembles of five or ten models, we extend our analysis to ensembles of two models to confirm our findings in resource-constrained environments.
    \item We release a replication package containing our experimental framework, raw data, and analysis scripts to encourage further research into LLM ensembles.\footref{foot:figshare-link}
\end{itemize}

\section{Related Work}

\citet{lu2024:merge} conducted a survey on collaborative strategies for LLMs. 
In total, they devised three collaboration types: merging, cooperation, and ensemble.
The \textit{merging} of LLMs is concerned with their parameters.  
As such, the parameters of multiple LLMs are 
fused
into a single model.
\textit{Cooperation} encompasses more complex types of interactions, such as the use of LLMs to check the factuality of outputs generated by another or the use of small LLMs to compress inputs for larger ones with the goal of improving efficiency. 
\textit{Ensemble} methods combine outputs generated by multiple LLMs. 
This can occur before, during, or after inference. 
Specifically, we are interested in LLM ensembles after inference, in which collaboration is achieved by selecting a subset of outputs generated from all LLMs.  

In addition to the survey by \citet{lu2024:merge}, \citet{ashiga2025:ensemble} surveyed ensemble methods for text and code generation with LLMs.
They devised seven types of LLM ensembles: weight merging, knowledge fusion, mixture-of-experts, reward ensemble, output ensemble, routing, and cascading.
Among these, the most popular approach for using ensembles was the use of mixture-of-experts~\cite{masoudnia2014:mixture}, where the output of a single LLM is determined from an architecture perspective. 
Similar to the mixture-of-agents approach, \citet{xue2024:multiprogramming} created ensembles with code generated in multiple languages from the same model (similar to MoE).
Other code-related tasks addressed with ensembles include vulnerability detection~\cite{ridoy2024:enstack,sun2025:ensembling, hort2025:semanticpreserving}, code search~\cite{du2021:single} and code generation~\cite{chen2025:debatecoder}.
Here, \citet{chen2025:debatecoder} employed two LLMs to iteratively generate code, and test cases for the respective other LLM.

The approach \textsc{LLM-BLENDER}, proposed by \citet{jiang2023:llmblender}, used ensembles of LLMs for instruction-based tasks and is able to consistently outperform single models. 
\textsc{LLM-BLENDER} consists of two components: PAIRRANKER and GENFUSER.
PAIRRANKER ranks outputs by performing a pairwise comparison between every generated output, to determine the better of the two candidates. 
For this purpose, metrics such as BERTScore~\cite{zhang2020:bertscore}, BLEURT~\cite{sellam2020:bleurt}, and BARTScore~\cite{yuan2021:bartscore} have been used to score each output according to its similarity to the ground truth. 
An alternative is the ranking of output pairs with ChatGPT, which did not require access to ground truth information. 
Unlike our goal to rank and choose multiple outputs from an ensemble of LLMs, PAIRRANKER ranks outputs, and GENFUSER uses this information to generate a single, final response. 

Other output ranking approaches, which can support the selection of outputs when dealing with an ensemble of LLMs, include the computation of cosine similarity between input and output~\cite{liu2021:simcls} or masked language modeling to compute the likelihood of generated outputs~\cite{salazar2020:masked}.
\citet{ravaut2023:summareranker} proposed SummaReranker, which learned a ranking model to determine the best solution from a set of candidates. 
For the task of program repair, confidence-based ranking is popular (i.e. how confident are the LLMs that the generated output solves the problem). 
As such, entropy has been used to rank the quality of patches~\cite{xia2023:automated,xia2022:less,yang2024:revisiting}.

While several approaches have been proposed for enabling the collaboration of LLMs, we notice a lack of research on outputs ensembles (i.e. selecting multiple outputs generated from an ensemble of LLMs), in particular when dealing with software engineering tasks.
Therefore, we set out to perform an empirical evaluation of output ensembles for two software engineering tasks (i.e. code generation and program repair). 

\section{Experiment Design}

\subsection{Research Questions}
We aim to answer the following research questions in our work:
\begin{enumerate}
    \item[RQ1] What are the performance differences between individual models vs. an ensemble of models for software engineering tasks?
    \begin{enumerate}
        \item[RQ1.1] To what degree do LLMs exhibit complementarity by solving unique sets of problems?
        \item[RQ1.2] How large is the performance gap between an ensemble's theoretical maximum and the best-performing individual model?
    \end{enumerate}
\end{enumerate}

To answer RQ1, we employ 10 different LLMs and 3 ensembles (Section~\ref{sec:design-models}).
We evaluate the performance of each individual model and each ensemble configuration on three benchmarks (Section \ref{sec:design-benchmarks}).
We first analyze model complementarity by studying the problem-solving capabilities within and across LLM families.
For each ensemble, we calculate the theoretical maximum score, which indicates the total number of problems solved by at least one model in a specific ensemble.
This quantifies the performance gap between a single model and the ensemble's theoretical maximum, establishing the ceiling for an ideal selection strategy.

\begin{enumerate}
    \item[RQ2] What heuristics are the most effective at achieving the theoretical potential of an LLM ensemble?
    \begin{enumerate}
        \item[RQ2.1] How do selection metrics and strategies compare in their ability to identify correct solutions from the pool of candidates?
        \item[RQ2.2] How consistent are the selection strategies when reducing the ensemble size to two models? %
    \end{enumerate}
\end{enumerate}

To address RQ2, we first implement and evaluate different heuristics to select outputs generated by ensembles of 5 or 10 models (Section \ref{sec:design-models}).
Each heuristic is composed of a selection metric (Section \ref{sec:design-select-metrics}) and a selection strategy (Section \ref{sec:design-strategies}).
We employ selection metrics that assign scores to each output based on either the confidence of the models or the similarity between outputs.
We then implement three selection strategies to favor consensus, disagreement, or diversity.
This allows us to compare each heuristic against the theoretical best possible score established in RQ1.
We then investigate the performance of these strategies on smaller ensembles composed of only 2 models.
We evaluate if the trends appearing on bigger ensembles persist and compare their performance against a Naive baseline, thereby assessing their utility in more resource-constrained environments.

\subsection{Models and Ensembles}
\label{sec:design-models}
To evaluate our research questions, we select a diverse set of 10 publicly available, instruction-tuned LLMs from five model families, all with coding capabilities.
For each family, we select two models of different parameter sizes.
For our analysis, we group these models into two categories: Small Models, which comprises models with approximately 7 to 8 billion parameters, and Large Models, which includes models in the 12 to 16 billion parameter range.
\autoref{tab:overview-models} provides an overview of the models in our study, along with abbreviations used 
for conciseness in Tables and Figures
throughout the paper. %
The abbreviations contain a subscript indicating if the model is the smaller variant (e.g. $MI_S$) or the larger variant (e.g. $MI_L$) in the family.

Based on these 10 models, we have created three ensembles.
The ensemble $Ens_{all}$ contains all ten models.
The ensemble $Ens_{S}$ contains the five small variants for each family.
Finally, the ensemble $Ens_{L}$ contains the five large variants for each family.

\begin{table}[t]
\vspace*{.5ex}
\centering
\caption{Overview of language models used.}
\label{tab:overview-models}

\begin{tabular}{@{}llrc}
\toprule
\textbf{Family} & \textbf{Model Name}                    & \textbf{Size} & \textbf{Abbr.}\\ \midrule
CodeLlama       & CodeLlama-7b\footref{foot:model-cl-s}                 & 7B            & $CL_S$                      \\
                & CodeLlama-13b\footref{foot:model-cl-l}                    & 13B           & $CL_L$                      \\ \midrule
DeepSeek        & DeepSeekCoder-6.7b\footref{foot:model-ds-s}              & 6.7B          & $DS_S$                      \\
                & DeepSeekCoder-V2-Lite\footref{foot:model-ds-l}           & 16B           & $DS_L$                      \\ \midrule
Gemma           & CodeGemma-7b\footref{foot:model-gm-s}                          & 7B            & $GM_S$                      \\
                & Gemma-3-12b\footref{foot:model-gm-l}                           & 12B           & $GM_L$                      \\ \midrule
Mistral         & Ministral-8B\footref{foot:model-mi-s}                & 8B            & $MI_S$                      \\
                & Mistral-Nemo\footref{foot:model-mi-l}                & 12B           & $MI_L$                      \\ \midrule
Qwen3           & Qwen3-8B\footref{foot:model-qw-s}                                  & 8B            & $QW_S$                      \\
                & Qwen3-14B\footref{foot:model-qw-l}                                & 14B           & $QW_L$                      \\ \bottomrule
\end{tabular}
\end{table}

\defnote{foot:model-cl-s}{\url{https://huggingface.co/codellama/CodeLlama-7b-Instruct-hf}}
\defnote{foot:model-cl-l}{\url{https://huggingface.co/codellama/CodeLlama-13b-Instruct-hf}}
\defnote{foot:model-ds-s}{\url{https://huggingface.co/deepseek-ai/deepseek-coder-6.7b-instruct}}
\defnote{foot:model-ds-l}{\url{https://huggingface.co/deepseek-ai/DeepSeek-Coder-V2-Lite-Instruct}}
\defnote{foot:model-gm-s}{\url{https://huggingface.co/google/codegemma-7b-it}}
\defnote{foot:model-gm-l}{\url{https://huggingface.co/google/gemma-3-12b-it}}
\defnote{foot:model-mi-s}{\url{https://huggingface.co/mistralai/Ministral-8B-Instruct-2410}}
\defnote{foot:model-mi-l}{\url{https://huggingface.co/mistralai/Mistral-Nemo-Instruct-2407}}
\defnote{foot:model-qw-s}{\url{https://huggingface.co/Qwen/Qwen3-8B}}
\defnote{foot:model-qw-l}{\url{https://huggingface.co/Qwen/Qwen3-14B}}

\subsection{Benchmarks}
\label{sec:design-benchmarks}
To ensure a comprehensive evaluation, we select three widely-used benchmarks representing two software engineering tasks: Automatic Program Repair (APR) and code generation. 
For the APR task, we use HumanEval-Java and Defects4J.
HumanEval-Java~\cite{jiang2023:impact} is a Java-based variant of the code synthesis benchmark.
This benchmark consists of 164 single-function bugs.
Defects4J v2.0~\cite{just2014:defects4j} comprises 835 bugs from large, open-source Java projects.
We follow the classification of previous work~\cite{xiang2024:how} and select the 525 single-function bugs.
Both APR benchmarks provide a potential fix along with unit tests to evaluate the plausibility of generated patches.

For the code generation task, we consider LiveCodeBench~\cite{jain2024:livecodebench}, a recent benchmark consisting of 1180 problems from competitive programming platforms compiled from May 2023 to May 2025.
From these, we select a subset of the latest 454 problems (from August 2024) set as default by the authors on their leaderboard.
This selection attempts to limit the data leakage on newer models.
Each problem consists of a problem description and may contain some starter code.
Unlike the previous benchmarks, LiveCodeBench does not contain any potential solution and relies only on a set of tests to assess plausibility.
By combining code-generation and program repair benchmarks, we aim to study the dynamics of LLM ensembles on different software engineering tasks and not constrain our analysis to a single one.

\subsection{Implementation}
The pipeline begins by building the initial prompt tailored to each benchmark.
We use a beam-based search decoding strategy with no stochastic sampling to keep all outputs deterministic and reproducible.
For the two APR benchmarks, the input consists of a function containing a bug.
The bug is delimited using the \texttt{<bug\_start>} and \texttt{<bug\_end>}.
The following template is used to generate the prompt:
\begin{verbatim}
"""
The input is buggy code. The bug starts from 
'<bug_start>' and ends at '<bug_end>'. 
Please fix the following code delimited by 
triple backticks, remove the '<bug_start>' 
and '<bug_end>' markers, and return the 
complete method fixed wrapped in triple backticks.
```java
{buggy_function}
```
"""
\end{verbatim}

On the other hand, the prompt for the code generation benchmark starts with a description of a problem which may include starting code.
We follow the template proposed in the original work~\cite{jain2024:livecodebench}:
\begin{verbatim}
"""
You are an expert Python programmer. You will be given a 
question (problem specification) and will generate a 
correct Python program that matches the specification 
and passes all tests. You will NOT return anything 
except for the program.

### Question: {question_content}

### Format: {formatting_message}

```python
{starter_code}
# YOUR CODE HERE
```
"""
\end{verbatim}

The formatting message for questions without starter code is:

\begin{verbatim}
    Read the inputs from stdin solve the problem and write 
    the answer to stdout (do not directly test on the 
    sample inputs). Enclose your code within delimiters 
    as follows.
\end{verbatim}

The formatting message for questions with started code is:

\begin{verbatim}
    You will use the following starter code to write 
    the solution to the problem and enclose your code 
    within delimiters.
\end{verbatim}

The outputs generated in either of the benchmarks are then parsed by extracting the blocks of text delimited by the triple backquote symbol (\texttt{\`{}\`{}\`{}}) that can be followed by a keyword (\texttt{\`{}\`{}\`{}java} or \texttt{\`{}\`{}\`{}python}).
Since models may output partial answers 
before or after the complete solution
for APR problems, if multiple blocks are delimited, we select the one that contains a full method declaration.

In the validation phase, we execute the code extracted (after being inserted in the appropriate context) and run the related tests.
The tests determine if the code is plausible or not.

\subsection{Family Advantage Index}
\label{sect:design-faiz}
We introduce the Family Advantage Index ($FAI_z$).
This 
z-score
quantifies how much more likely a large model is to solve a hard problem that its smaller counterpart solved, relative to the average performance of other large models.
We define a hard problem as one not solved by all small models, ensuring that very easy problems solved by everyone are excluded from the calculation.
A positive $FAI_z$ score indicates the degree of family advantage, i.e. the larger model is more likely to solve the same hard problems that its smaller counterpart solved compared to other larger models.
However, it is critical to acknowledge potential limitations of this score.
$FAI_z$ is sensitive to performance variations of other models used to calculate the average, and to the definition of what is considered a hard problem.

Formally, let $\mathcal{H}$ be the set of hard problems (those solved by at least one but not all small models in our study). 
Given a pair of models $(M_S, M_L)$ from the same family, we calculate the conditional probability that $M_L$ solves a hard problem, given $M_S$ solved it: $p_L = P(M_L \mid M_S, \mathcal{H})$. 
We then compute the mean ($\mu$) and standard deviation ($\sigma$) of this same conditional probability for all other large models in our set. 
The Family Advantage Index is the resulting z-score:
\begin{equation}
    FAI_z(M_S \to M_L) = \frac{p_L - \mu}{\sigma}
\end{equation}

\subsection{Selection Metrics}
\label{sec:design-select-metrics}

The metrics chosen can be divided into two categories: output-based metrics and confidence-based metrics.
\textbf{Output-based} metrics measure semantic and syntactic properties between two pieces of code.
We select two output-based metrics: CodeBERTScore~\cite{zhou2023:codebertscore} and CodeBLEU~\cite{ren2020:codebleu}.
CodeBERTScore uses the embeddings from the CodeBERT model~\cite{feng2020:codebert} to compute the cosine similarity between two pieces of code.
We choose CodeBERTScore F3 since it is recommended by the authors for functional correctness.
CodeBLEU measures syntactic similarity by adapting BLEU~\cite{papineni2002:bleu} to source code.
It calculates a weighted combination of the original BLEU score, n-gram match, abstract syntax tree match, and data-flow match.
For both metrics, we compute the final score of a candidate by summing over all scores of said candidate against the rest of the candidates for the same problem.

On the other hand, \textbf{confidence-based} metrics employ a model's internal confidence to assess an output.
Prior work~\cite{xia2023:automated,yang2024:revisiting} has successfully used confidence-based metrics to rank patches.
We select three confidence-based metrics: NLL / Byte, Entropy / Byte, and Normalized Sum Entropy.

Let $t_{1:n}$ be the tokens of a patch, and let $p_i$ denote the model probability assigned to token $t_i$ given the preceding tokens. 
We define negative log-likelihood (NLL) and the per-position predictive entropy 
as:
\begin{equation}
\mathrm{NLL} \;=\; -\sum_{i=1}^{n} \log p_i, 
\qquad
E_i \;=\; -\sum_{v\in\mathcal{V}} q_i(v)\,\log q_i(v),
\end{equation}
where $q_i(\cdot)$ is the model's 
distribution at position $i$, $\mathcal{V}$ is the vocabulary, and all logarithms are natural (in nats). 
Let $m$ be the number of bytes (in UTF-8) in the generated patch, and let $\log|\mathcal{V}|$ be the maximum entropy of a uniform distribution over the vocabulary. We then compute: 

\begin{align}
\texttt{nll\_per\_byte} 
&= \frac{\mathrm{NLL}}{m} 
, \label{eq:nll_per_byte}\\[2pt]
\texttt{entropy\_per\_byte} 
&= \frac{\sum_{i=1}^{n} E_i}{m}, \label{eq:entropy_per_byte}\\[2pt]
\texttt{sum\_entropy\_norm} 
&= \frac{\sum_{i=1}^{n} E_i}{\log|\mathcal{V}|}. \label{eq:sum_entropy_norm}
\end{align}

The first two metrics are normalized by bytes to mitigate tokenization effects and enable comparison across models and tokenizers.
This type of normalization cannot be applied to the entropy sum.
However, we still employ this metric inspired by \citet{xia2023:automated} who obtained better outcomes favoring shorter and simpler code following Occam’s razor hypothesis~\cite{sober2015:ockhams}.
To reduce the effects of vocabulary differences between the models on the entropy sum, we normalize according to the vocabulary size of each tokenizer.
For these 3 metrics, we calculate the final score of a candidate by summing over all confidence-based scores of said candidate given by all the models in the ensemble.

\subsection{Candidate Pool Construction}
For every problem in our benchmarks, we prompt each of the 10 models to generate $n=10$ outputs.
The outputs from all models are aggregated in a large pool of candidates.
From this set, we select a subset of $k=10$ outputs.
This constraint is motivated by two practical considerations.
First, developers are unlikely to review more than 10 patches~\cite{noller2022:trust}.
Second, the computational cost of validating outputs can be high, especially for large-scale projects found in benchmarks such as Defects4J~\cite{liu2020:efficiency}.
These decisions align with common practices balancing efficient use of resources with the need for a diverse candidate pool, and allow for comparison with previous work~\cite{jiang2023:impact,silva2025:repairllama, ruiz2025:art}.

\subsection{Selection Strategies}
\label{sec:design-strategies}
Given a set of scored candidates, we investigate three strategies to choose the final subset of $k=10$ candidates: Highest, Lowest, and Diversity.
The first two strategies select the 10 outputs with the highest or lowest scores.
The goal of these strategies is to select patches based on consensus and disagreement.
The last strategy aims to maximize diversity within the set of candidate outputs.
We employ a greedy approach starting with the two candidates with the highest and lowest scores. 
It then iteratively builds a set of 10 candidates by selecting candidates that maximize the distance to candidates already selected.
The goal is to explore a wider range of potential solutions rather than focusing on minor variations of a single popular but incorrect candidate.

To establish a baseline, we also employ a Naive selection strategy that does not take into account any selection metric.
This Naive strategy allows each model to contribute equally. 
Since we restrict the number of outputs per ensemble to $k=10$, we build the selection with the first 1 or 2 outputs per model in ensembles of 10 or 5 models respectively.

\subsection{Evaluation Metric}
We assess the effectiveness of the different ensembles and strategies by measuring the number of problems in each benchmark with at least one plausible candidate.
A candidate solution is considered plausible if it compiles successfully and passes the test suite associated with the problem.
The candidate is considered implausible otherwise.
Given that our experiments use 10 different models generating 10 outputs per problem, the test suite associated with each problem allows us to efficiently evaluate tens of thousands of candidate solutions.

\section{Results and Discussion}
\subsection{Results of RQ1}
\subsubsection{Complementarity of LLMs}

\begin{table}[t]
\vspace*{.5ex}
\caption{[RQ1.1] Problems solved ($M_S\to M_L$) and corresponding \( \mathrm{FAI}_z \) per family and benchmark. Each cell shows the number of problems solved for the smaller model $M_S$ and larger model $M_L$; parentheses contain \( \mathrm{FAI}_z \) computed on the benchmark's set of hard problems.}
\label{tab:solved_faiz}
\centering
\begingroup
\resizebox{\columnwidth}{!}{
\begin{tabular}{@{}l@{}rr@{}r@{}}
\toprule
\textbf{Family} & \multicolumn{1}{@{}c}{\textbf{HumanEval-Java}} & \multicolumn{1}{c}{\textbf{Defects4J}} & \multicolumn{1}{r@{}}{\textbf{LiveCodeBench}} \\
\midrule
CodeLlama & 59$\to$\hphantom{1}64$^{(-6.35)}$ & 42$\to$\hphantom{1}47$^{(+0.00)}$  & 56$\to$\hphantom{1}55$^{(-8.49)}$ \\
DeepSeek  & \textbf{110}$\to$\textbf{122}$^{(+1.02)}$& \textbf{89}$\to$\textbf{112}$^{(+1.11)}$ & 87$\to$110$^{(+0.04)}$ \\
Gemma     & 101$\to$108$^{(+0.56)}$& 84$\to$\hphantom{1}87$^{(-0.46)}$ & 80$\to$122$^{(+1.31)}$ \\
Mistral   & 108$\to$104$^{(+0.21)}$& 77$\to$\hphantom{1}81$^{(+0.40)}$  & 87$\to$\hphantom{1}92$^{(+0.02)}$ \\
Qwen      & 98$\to$120$^{(+0.88)}$ & 65$\to$111$^{(+1.90)}$ & \textbf{130}$\to$\textbf{155}$^{(+2.20)}$ \\
\bottomrule
\end{tabular}
}
\endgroup
\end{table}

\begin{figure}[t]
\vspace*{.5ex}
    \begin{subfigure}[b]{0.45\textwidth}
        \includegraphics[width=\linewidth]{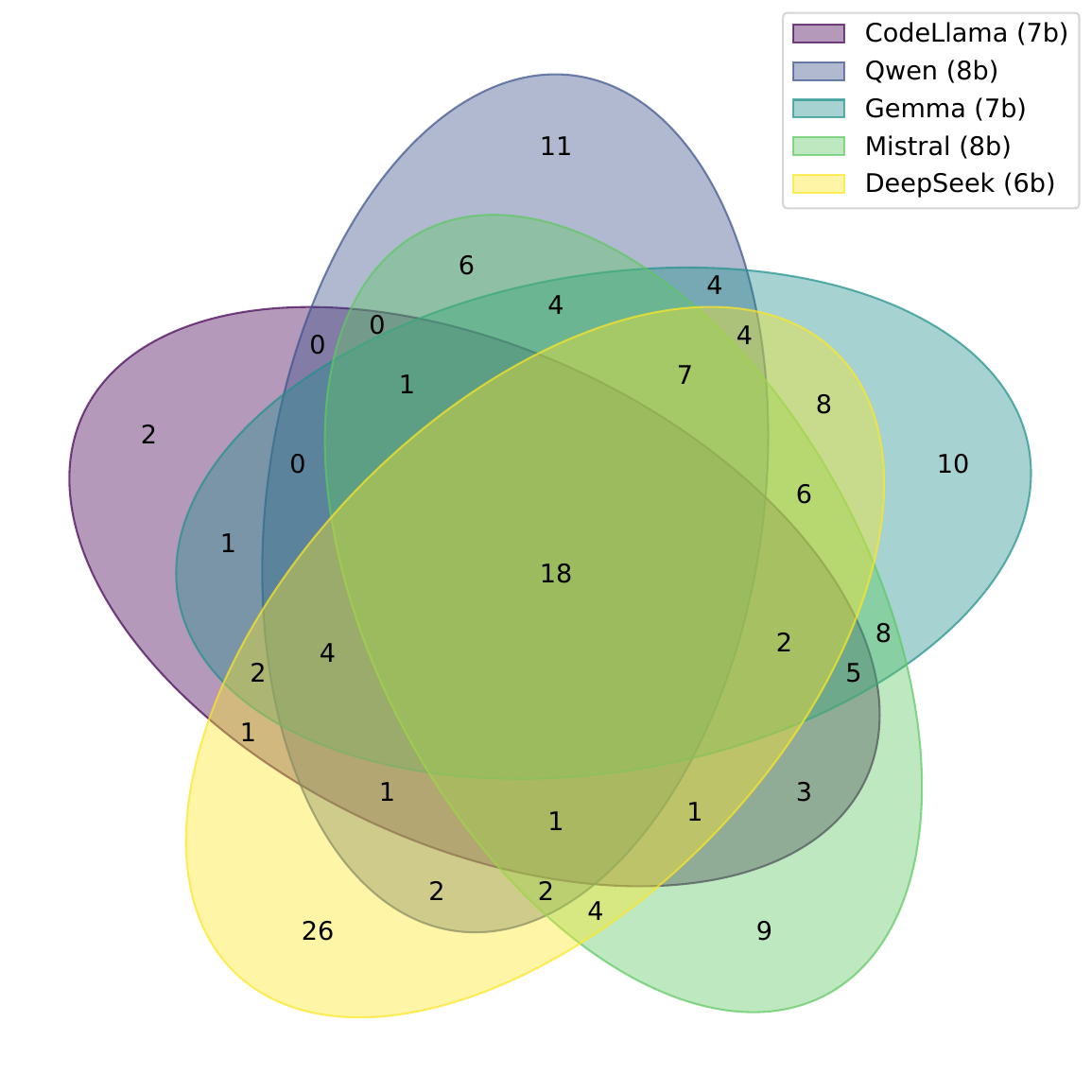}
        \subcaption{Small Models}\label{fig:venn-d4j-small}
    \end{subfigure}
    \begin{subfigure}[b]{0.45\textwidth}
        \includegraphics[width=\linewidth]{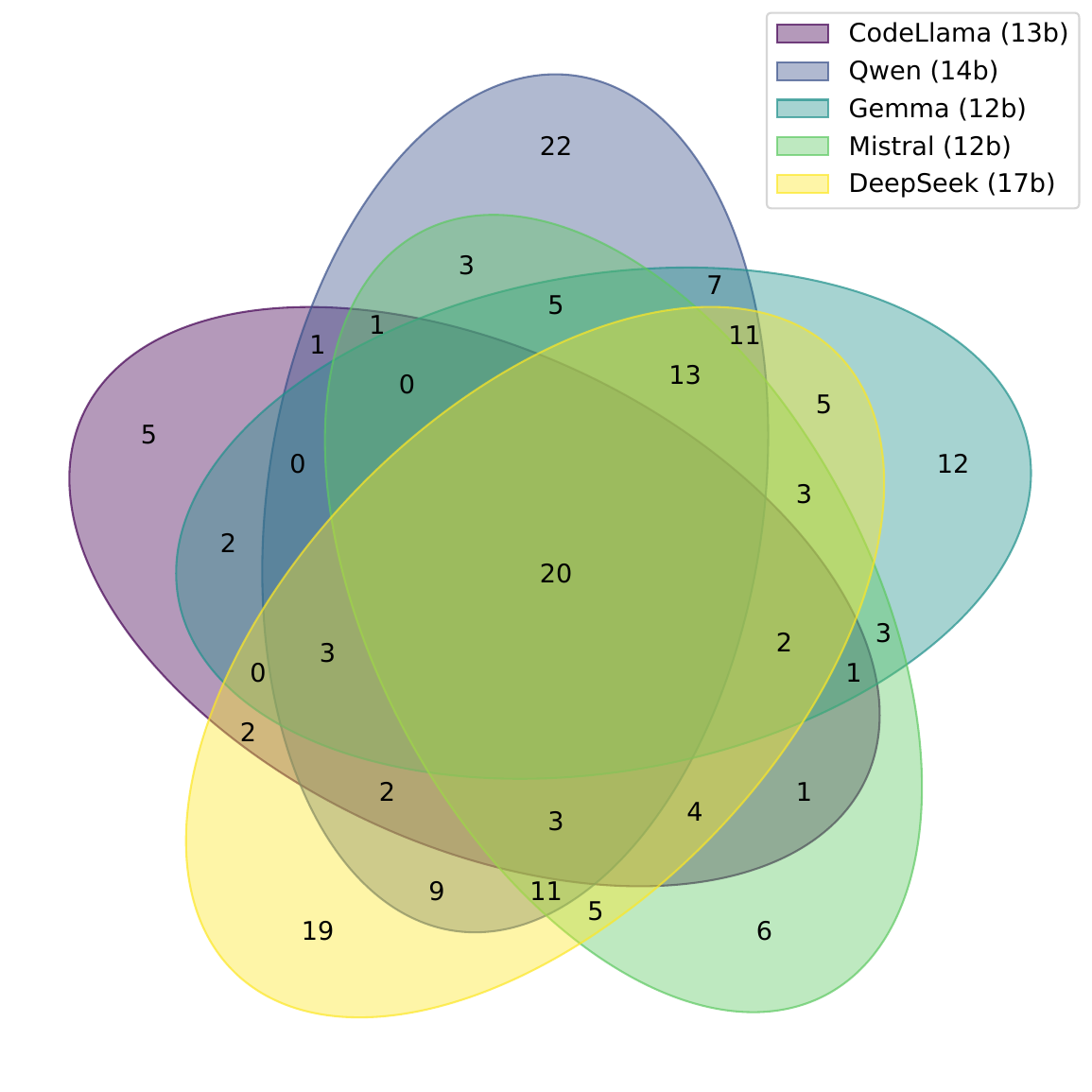}
        \subcaption{Large Models}\label{fig:venn-d4j-big}
    \end{subfigure}
    \caption{[RQ1.1] Venn diagram of problems solved for Defects4J benchmark for two sets of models indicating their complementarity.}
    \label{fig:general-venn-diag-llama31}
    \Description[Short Descr]{Long Descr}
\end{figure}

An ensemble is expected to achieve better problem-solving capabilities than its best individual member.
Thus, we need to analyze the ensemble's theoretical performance ceiling before constructing it.
To this end, we start by studying the relationship between the small ($M_S$) and large ($M_L$) variants of the same family before diving into bigger ensembles.

A common assumption is that larger models solve at least all the problems their smaller counterparts can~\cite{kaplan2020:scaling, hassid2024:larger}.
Our results quickly prove this assumption incorrect, as shown by the number of problems solved for each model in \autoref{tab:solved_faiz}.
For instance, CodeLlama (13b) solves fewer problems than CodeLlama (7b) on LiveCodeBench, proving it cannot be a superset.
Another instance occurs in the Mistral family, where Mistral (8b) solves more problems on HumanEval-Java than Mistral (12b).
This demonstrates that smaller models are still relevant and are not made redundant by larger versions.

Despite this complementarity, the results also indicate that larger models tend to solve more problems overall.
Therefore, we first study these intra-family relationships quantitatively.

To this end, we use the Family Advantage Index ($FAI_z$) described in Section \ref{sect:design-faiz}, a z-score that measures how many standard deviations a large model's success rate on hard problems deviates from the average success rate of other large models on the same problems.

The number of problems solved for each model and the $FAI_z$ for each family are shown in \autoref{tab:solved_faiz}.
Our results show that a family advantage often 
exists, but also exhibits high variance.
The Qwen family exhibits a strong positive $FAI_z$ on Defects4J ($+1.90$) and LiveCodeBench ($+2.20$).
In contrast, the CodeLlama family shows an even stronger negative $FAI_z$ on HumanEval-Java ($-6.35$) and LiveCodeBench ($-8.49$).
These results indicate that even within the same family, individual model variants may offer distinct solutions.
However, to increase the likelihood that an ensemble has the best chances to achieve higher outputs, including models from different families is preferred.

To confirm the viability of large-scale ensembles, we analyze the unique contribution of individual models to an ensemble.
The existence of problems solved by only a single model, including models with a lower overall score, indicates complementarity between the models.
We illustrate this complementarity through Venn diagrams composed of models of similar sizes.
For brevity, we include only two out of nine diagrams, the remaining ones can be found in our public repository.\footref{foot:figshare-link}
\autoref{fig:venn-d4j-small} include smaller models while \autoref{fig:venn-d4j-big} encompass their larger counterparts applied to Defects4J.

Even among the smaller models, the set of unique problems solved is substantial.
DeepSeek (6B), the highest score in this benchmark, solves 26 unique bugs that no other small model could solve.
Gemma (7b) solves only 5 problems fewer than DeepSeek (6b), yet contributes solutions for 10 unique bugs.
While a model like CodeLlama (7b) solves fewer than half the number of problems of DeepSeek (6B), it is still able to provide solutions to two unique problems.

This trend repeats itself in the case of larger models.
Lower-performing models like CodeLlama (13b) provide 5 unique solutions despite maintaining an overall score lower than half of the best model.
In this case, the model with the most unique solution is Qwen (14b) with 22 unique bugs solved, followed by DeepSeek (16b) with 19 unique bugs solved.

\begin{table}[t]
\vspace*{.5ex}
\caption{[RQ1.2] Comparison of the number of problems solved by the single best small ($M_S$) and large ($M_L$) model against the theoretical maximum number of problems solved by 
an ensemble of five small models, five large models, and all ten models combined.
}
\label{tab:theoretical-max-gap}
\centering
\begin{tabular}{@{}lcccc@{}}
\toprule
\multirow[b]{2}{*}{\makecell[tc]{\textbf{Benchmark}}} &
\multirow[t]{2}{*}{\makecell[tc]{\textbf{Best single}\\\textbf{model }(\#)\\\scriptsize($M_S \to M_L$)}} &
\multicolumn{3}{c}{\makecell[tc]{\textbf{Theoretical}\\\textbf{max} (\#)}} \\
\cline{3-5}
\addlinespace[2pt]
& & \textbf{$Ens_S$} & \textbf{$Ens_L$} & \textbf{$Ens_{all}$} \\
\midrule
HumanEval-Java & 110$\to$122 & 132 & 139 & 141 \\
Defects4J & \hphantom{1}89$\to$112 & 153 & 181 & 205 \\
LiveCodeBench & 130$\to$155 & 154 & 177 & 185 \\
\bottomrule
\end{tabular}
\end{table}

\begin{table*}[t]
\vspace*{.5ex}
\centering
\caption{[RQ2.1] Problems solved HumanEval-Java. Bolding shows the best option per ensemble.}
\label{tab:results-all-hej-modified}
{%
\setlength{\aboverulesep}{0pt}%
\setlength{\belowrulesep}{0pt}%
\resizebox{0.9\textwidth}{!}{
\begin{tabular}{l
  >{\columncolor{lightgray}}c >{\columncolor{lightgray}}c >{\columncolor{lightgray}}c
  c c c
  >{\columncolor{lightgray}}c >{\columncolor{lightgray}}c >{\columncolor{lightgray}}c
  c c c
  >{\columncolor{lightgray}}c >{\columncolor{lightgray}}c >{\columncolor{lightgray}}c
  c >{\columncolor{lightgray}}c}
\toprule
\multirow{2}{*}{\textbf{Ensemble}}
  & \multicolumn{3}{c}{\cellcolor{lightgray}\textbf{CodeBERT F3}}
  & \multicolumn{3}{c}{\textbf{CodeBLEU}}
  & \multicolumn{3}{c}{\cellcolor{lightgray}\textbf{NLL/Byte}}
  & \multicolumn{3}{c}{\textbf{Entropy/Byte}}
  & \multicolumn{3}{c}{\cellcolor{lightgray}\textbf{Sum Entropy}}
  & \multirow{2}{*}{\textbf{Naive}} & \multirow{2}{*}{\cellcolor{lightgray}} \\
\cmidrule(lr){2-4} \cmidrule(lr){5-7} \cmidrule(lr){8-10} \cmidrule(lr){11-13} \cmidrule(lr){14-16}
& \textbf{H} & \textbf{L} & \textbf{D}
& \textbf{H} & \textbf{L} & \textbf{D}
& \textbf{H} & \textbf{L} & \textbf{D}
& \textbf{H} & \textbf{L} & \textbf{D}
& \textbf{H} & \textbf{L} & \textbf{D}
& & \multirow{-2}{*}{\cellcolor{lightgray}\textbf{Best}}\\
\midrule
$Ens_{S}$ & 64 & 108 & \textbf{129} & 64 & 110 & 128 & 75 & 109 & 125 & 75 & 111 & 125 & 103 & 90 & 125 & 112 & 132\\
$Ens_{L}$ & 83 & 118 & 130 & 82 & 124 & 131 & 94 & 122 & \textbf{135} & 95 & 123 & 134 & 117 & 94 & 128 & 127 & 139\\
$Ens_{all}$   & 69 & 111 & 130 & 69 & 109 & 127 & 79 & 114 & \textbf{133} & 78 & 113 & 132 & 110 & 91 & 130 & 123 & 141\\
\bottomrule
\multicolumn{18}{@{}c@{}}{\footnotesize \textit{H / L / D} = Highest / Lowest / Diversity} \\
\end{tabular}
}
}%
\end{table*}

\begin{table*}[t]
\centering
\caption{[RQ2.1] Problems solved Defects4J. Bolding shows the best option per ensemble.}
\label{tab:results-all-d4j-modified}
{%
\setlength{\aboverulesep}{0pt}%
\setlength{\belowrulesep}{0pt}%
\resizebox{0.9\textwidth}{!}{
\begin{tabular}{l
  >{\columncolor{lightgray}}c >{\columncolor{lightgray}}c >{\columncolor{lightgray}}c
  c c c
  >{\columncolor{lightgray}}c >{\columncolor{lightgray}}c >{\columncolor{lightgray}}c
  c c c
  >{\columncolor{lightgray}}c >{\columncolor{lightgray}}c >{\columncolor{lightgray}}c
  c >{\columncolor{lightgray}}c}
\toprule
\multirow{2}{*}{\textbf{Ensemble}}
  & \multicolumn{3}{c}{\cellcolor{lightgray}\textbf{CodeBERT F3}}
  & \multicolumn{3}{c}{\textbf{CodeBLEU}}
  & \multicolumn{3}{c}{\cellcolor{lightgray}\textbf{NLL/Byte}}
  & \multicolumn{3}{c}{\textbf{Entropy/Byte}}
  & \multicolumn{3}{c}{\cellcolor{lightgray}\textbf{Sum Entropy}}
  & \multirow{2}{*}{\textbf{Naive}} & \multirow{2}{*}{\cellcolor{lightgray}} \\
\cmidrule(lr){2-4} \cmidrule(lr){5-7} \cmidrule(lr){8-10} \cmidrule(lr){11-13} \cmidrule(lr){14-16}
& \textbf{H} & \textbf{L} & \textbf{D}
& \textbf{H} & \textbf{L} & \textbf{D}
& \textbf{H} & \textbf{L} & \textbf{D}
& \textbf{H} & \textbf{L} & \textbf{D}
& \textbf{H} & \textbf{L} & \textbf{D}
& & \multirow{-2}{*}{\cellcolor{lightgray}\textbf{Best}}\\
\midrule
$Ens_{S}$ & 26 & 114 & 134 & 22 & 118 & 135 & 77 & \hphantom{0}80 & 133 & 75 & \hphantom{0}93 & \textbf{137} & 100 & 54 & 128 & \hphantom{0}88 & 153\\
$Ens_{L}$ & 40 & 134 & 153 & 35 & 137 & \textbf{162} & 86 & 115 & 152 & 81 & 123 & 154 & 135 & 53 & 157 & 116 & 181\\
$Ens_{all}$   & 22 & 125 & 157 & 25 & 129 & 164 & 87 & 116 & 152 & 86 & 122 & 161 & 133 & 53 & \textbf{167} & \hphantom{0}97 & 205\\
\bottomrule
\multicolumn{18}{@{}c@{}}{\footnotesize \textit{H / L / D} = Highest / Lowest / Diversity} \\
\end{tabular}
}
}%
\end{table*}

\begin{table*}[t]
\centering
\caption{[RQ2.1] Problems solved LiveCodeBench. Bolding shows the best option per ensemble.}
\label{tab:results-all-lcb-modified}
{%
\setlength{\aboverulesep}{0pt}%
\setlength{\belowrulesep}{0pt}%
\resizebox{0.9\textwidth}{!}{
\begin{tabular}{l
  >{\columncolor{lightgray}}c >{\columncolor{lightgray}}c >{\columncolor{lightgray}}c
  c c c
  >{\columncolor{lightgray}}c >{\columncolor{lightgray}}c >{\columncolor{lightgray}}c
  c c c
  >{\columncolor{lightgray}}c >{\columncolor{lightgray}}c >{\columncolor{lightgray}}c
  c >{\columncolor{lightgray}}c}
\toprule
\multirow{2}{*}{\textbf{Ensemble}}
  & \multicolumn{3}{c}{\cellcolor{lightgray}\textbf{CodeBERT F3}}
  & \multicolumn{3}{c}{\textbf{CodeBLEU}}
  & \multicolumn{3}{c}{\cellcolor{lightgray}\textbf{NLL/Byte}}
  & \multicolumn{3}{c}{\textbf{Entropy/Byte}}
  & \multicolumn{3}{c}{\cellcolor{lightgray}\textbf{Sum Entropy}}
  & \multirow{2}{*}{\textbf{Naive}} & \multirow{2}{*}{\cellcolor{lightgray}} \\
\cmidrule(lr){2-4} \cmidrule(lr){5-7} \cmidrule(lr){8-10} \cmidrule(lr){11-13} \cmidrule(lr){14-16}
& \textbf{H} & \textbf{L} & \textbf{D}
& \textbf{H} & \textbf{L} & \textbf{D}
& \textbf{H} & \textbf{L} & \textbf{D}
& \textbf{H} & \textbf{L} & \textbf{D}
& \textbf{H} & \textbf{L} & \textbf{D}
& & \multirow{-2}{*}{\cellcolor{lightgray}\textbf{Best}}\\
\midrule
$Ens_{S}$ & 102 & 103 & \textbf{143} & 102 & 101 & 141 & 99 & 106 & 140 & 99 & 105 & 138 & 103 & 107 & 138 & 142 & 154\\
$Ens_{L}$ & 110 & 130 & 167 & 103 & 138 & \textbf{170} & 91 & 126 & 167 & 91 & 125 & 167 & 108 & 130 & 168 & 168 & 177\\
$Ens_{all}$ & 104 & 110 & 162 & 103 & 107 & 160 & 89 & 116 & 158 & 89 & 112 & 159 & \hphantom{0}96  & 125 & \textbf{170} & 169 & 185\\
\bottomrule
\multicolumn{18}{@{}c@{}}{\footnotesize \textit{H / L / D} = Highest / Lowest / Diversity} \\
\end{tabular}
}
}%
\end{table*} 
\subsubsection{Performance gap between ensembles and individual models}
The complementarity shown in the previous experiments indicates the existence of a performance ceiling.
We define the theoretical maximum as the total number of unique problems solved by at least one model within a set of models, representing the best possible outcome under a perfect selection approach.
The performance gap between the best single model of an ensemble to this theoretical maximum is substantial. 

\autoref{tab:theoretical-max-gap} presents this analysis across our three benchmarks.
The greatest gap in performance occurs on the Defects4J benchmark.
While the best small model solves 89 problems, an ensemble of small models could reach 153 fixed problems.
Similarly, the best large model solves 112 problems, while an ensemble of large models can solve 181 problems.
Comparing
against the ensemble of all the models, the results indicate a potential performance improvement of 83\% over the best individual model.

While the gap is largest for the Defects4J benchmark, it remains substantial across all benchmarks.
Both benchmarks, LiveCodeBench and HumanEval-Java, also see an improvement between $10\%$ to $20\%$ between the highest score in an ensemble of models to the theoretical maximum score achieved by the ensemble.
Furthermore, the data show that an ensemble of all models consistently outperforms ensembles of large and small models.
For example, on Defects4J, the larger model ensemble could solve 181 problems, but adding the five smaller models increases this number by 24 unique solutions (a $13\%$ increase).
This finding further validates our earlier observations through the $FAI_z$ analysis, confirming that smaller models do not always produce subsets of solutions of their larger counterparts.

Our previous experiments have established that LLM ensembles have substantial theoretical potential.
For instance, the best model on the Defects4J benchmark, DeepSeek (16b), repairs 112 bugs while the ensemble of all models collectively could increase this number to 205, indicating a performance increase of 83\%.
However, this represents an idealized upper bound achievable only with the perfect selection oracle.
Therefore, there is a need to study different selection mechanisms capable of selecting correct solutions from the pool of candidates generated by an ensemble.

\subsection{Results of RQ2}

Given a pool of outputs generated by the models of an ensemble, the challenge lies in selecting the most likely candidate outputs for validation, since testing all pool of outputs is often infeasible.
We start this analysis by first establishing a baseline with a Naive strategy.
The results from this approach are shown under the column \emph{Naive} in \autoref{tab:results-all-hej-modified}, \autoref{tab:results-all-d4j-modified}, and \autoref{tab:results-all-lcb-modified}.
The performance of this Naive heuristic varies across the benchmarks.
We obtain the best results on LiveCodeBench, for instance, 
the ensemble of small models
solved 142 problems, 
improving the best single model's score (130), but still far from reaching the theoretical maximum of 185.
This gap in performance decreases on HumanEval-Java, where 
the ensemble of large models
solves 127 problems contrasting with the 122 problems solved by the best individual model.
But, still far from the theoretical maximum of 139, while for other ensembles the results barely improve the performance of the best model.
For the more complex Defects4J benchmark, this strategy improves the score only for 
the ensemble of large models
while decreasing it for the other two ensembles when compared to the performance of the best model.
These results underscore the need for more sophisticated selection mechanisms.

\subsubsection{Outputs-based Selection Strategies}
We first focus on output-based metrics, which assess the similarity of an output to other candidates in the pool.
The results are detailed in \autoref{tab:results-all-hej-modified}, \autoref{tab:results-all-d4j-modified}, and \autoref{tab:results-all-lcb-modified}.
A key finding is the consistent failure of consensus-based selection.
Across all benchmarks and ensembles, selecting the candidates with the highest CodeBLEU and CodeBERT scores results in poor performance, worse even than the 
Naive baseline approach.
This phenomenon, which we term the \textit{popularity trap}, suggests that models frequently produce syntactically similar but semantically incorrect solutions.
In other words, models frequently fail in the same manner across their generated outputs.
Relying on model consensus amplifies this phenomenon, often filtering out correct solutions for problems that not all models could solve.

Strategies based on disagreement (lowest scores) improve the performance substantially, as they avoid the popularity trap.
However, this approach consistently outperforms the Naive heuristic only on the Defects4J benchmark.
By relying on the lowest similarity, 
we are selecting any output with a hallucination that makes it significantly different from the rest of the generated outputs.
Looking at the previous two approaches, we can agree that ensembles would benefit from a heuristic that does not fall into the popularity trap but neither allows arbitrary hallucinations.
Therefore, we arrive at the third proposed strategy based on diversity.

The results clearly indicate that the diversity approach consistently improves the performance over the previous two approaches across both metrics and all three benchmarks.
This heuristic proves itself most beneficial on the Defects4J benchmark, where the ensemble of all models selects a correct output for 164 out of 205 theoretically solvable problems.
This represents 80\% of its theoretical potential, substantially exceeding the 97 correct outputs achieved by the Naive approach.
On the other hand, on HumanEval-Java, the diversity strategy ranges from slightly improving the Naive approach by 3 on the ensemble of large models, to more substantial benefits on the ensemble of small models, where it realizes over 95\% of the ensemble's theoretical potential.
Finally, in LiveCodeBench, the diversity strategy achieves approximately the same number of problems solved except for the ensembles of all models, where it decreases the number up to 9 compared to the Naive approach.

\subsubsection{Confidence-based Selection Strategies}
In contrast to output-based metrics, confidence-based metrics follow a more expected trend.
For the three confidence-based metrics, lower values indicate that the output is more natural or that a model is more confident in predicting it.
Thus, selecting outputs with lower scores indicates consensus between the models.
We observe similar trends for NLL/Byte and Entropy/Byte, where consensus consistently achieves better scores than disagreement (selecting the highest) across benchmarks and ensembles.
We suspect that using the model's internal confidence scores helps avoid the popularity trap described in the previous subsection.
However, Sum Entropy surprisingly does not follow the same trend for the two APR benchmarks.
While the best scores on the HumanEval-Java benchmark are lower compared to the other two metrics, we notice a substantial difference on the Defects4J benchmark.
Specifically, in the latter benchmark, the best score from a consensus strategy is 123 problems solved 
by the ensemble of large models
using Entropy/Byte, contrasting with the disagreement approach with Sum Entropy which solves 135 problems.
Since this metric is length-dependent, selecting the highest scores can function similarly to selecting the lowest scores in similarity output-based metrics.
This would explain the similar scores between the two approaches, especially the sudden increase in LiveCodeBench 
in the ensemble of large models.
Therefore, although achieving higher scores than other confidence-based metrics, it suffers from similar limitations where an output containing a longer hallucination from the model would receive a higher score.

Similar to the output-based metrics, these confidence-based metrics benefit from a diversity heuristic across all benchmarks.
On the Defects4J benchmark, the best score of 135 problems achieved with the previous strategies is increased to 157 by switching to this heuristic.
Furthermore, on this benchmark, a new best score of 167 solved problems is achieved by increasing the size of the ensemble to all models.
Diversity improves the best results across benchmarks, increasing from 123 to 135 solved problems in HumanEval-Java or improving from 130 to 170 problems solved in LiveCodeBench.

\begin{figure*}[t]
    \begin{subfigure}[b]{0.3\textwidth}
        \includegraphics[width=\linewidth]{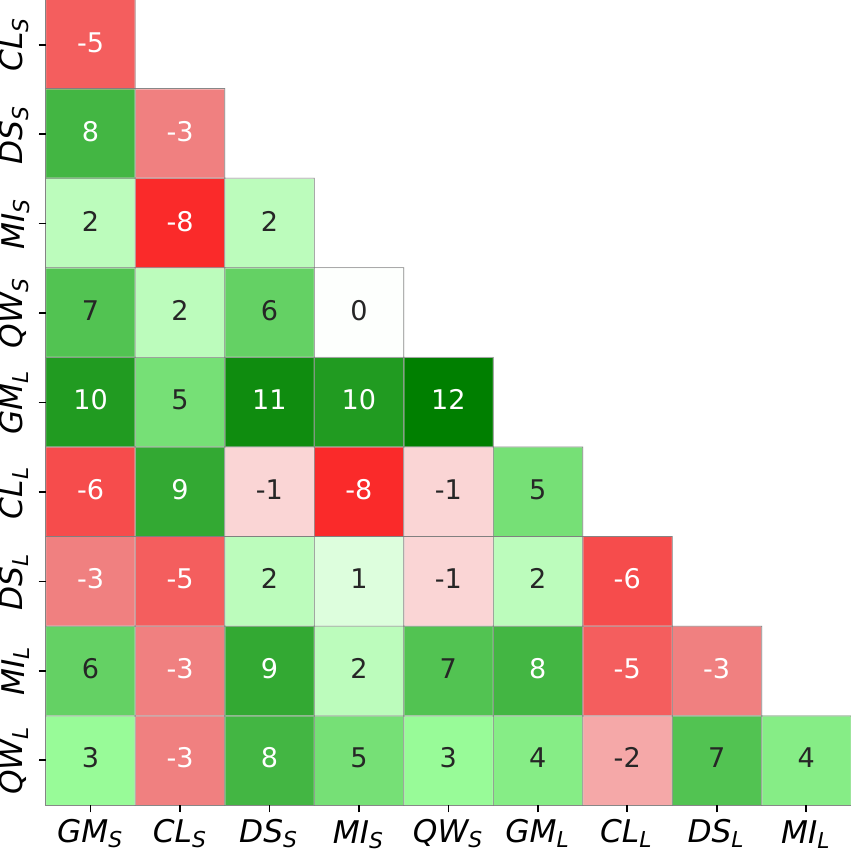}
        \subcaption{HumanEval-Java}\label{fig:heatmap-hej-Naive}
    \end{subfigure}
    \begin{subfigure}[b]{0.3\textwidth}
        \includegraphics[width=\linewidth]{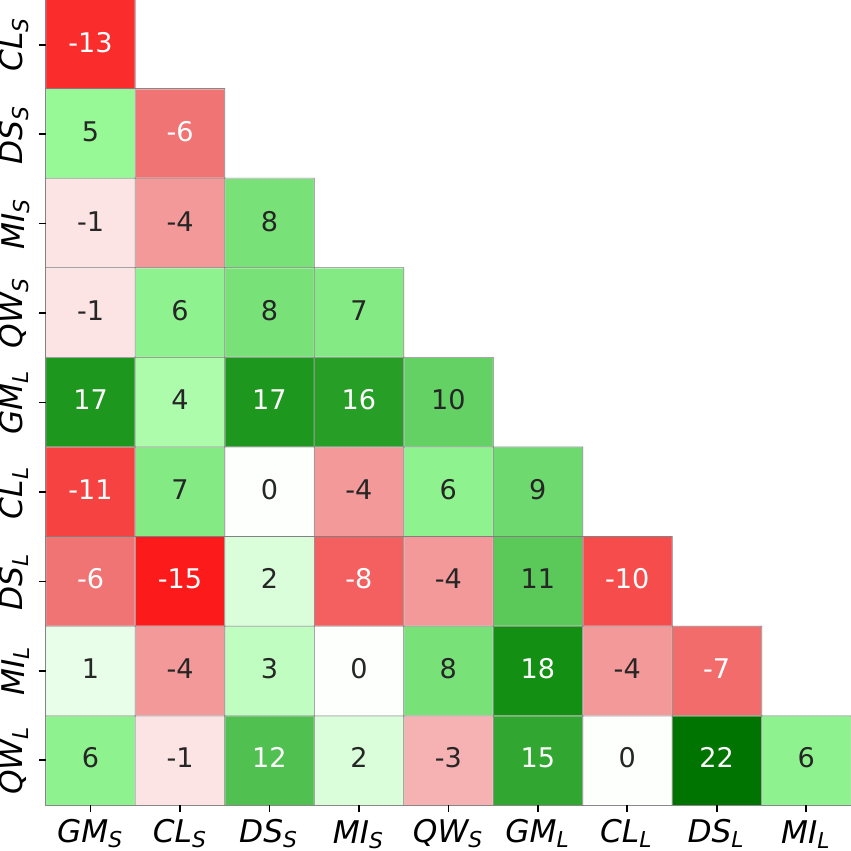}
        \subcaption{Defects4J}\label{fig:heatmap-d4j-Naive}
    \end{subfigure}
    \begin{subfigure}[b]{0.3\textwidth}
        \includegraphics[width=\linewidth]{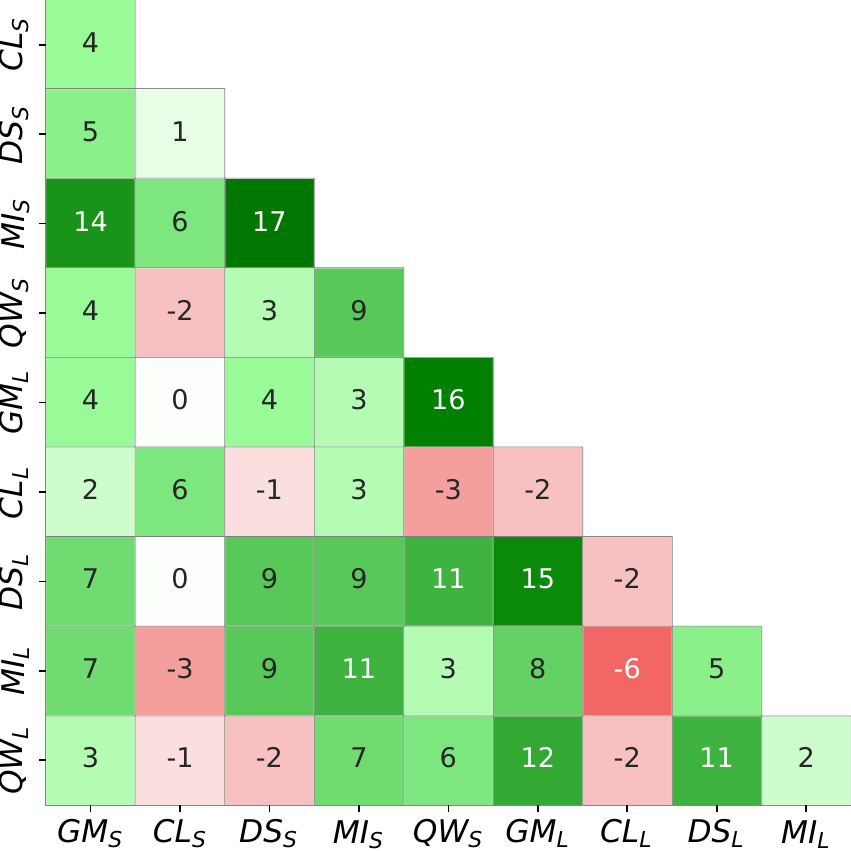}
        \subcaption{LiveCodeBench}\label{fig:heatmap-lcb-Naive}
    \end{subfigure}
    \caption{[RQ2.2] Comparison of two-model ensembles with Naive strategy vs. the best score of the two models. The heatmaps show the difference in the number of plausible candidates per benchmark (e.g. a negative score indicates that the Naive ensemble strategy performs worse than picking the better single model).}
    \label{fig:heatmaps-general-Naive}
    \Description[Short Descr]{Long Descr}
\end{figure*}

\begin{figure*}[t]
\vspace*{1ex}
    \begin{subfigure}[b]{0.3\textwidth}
        \includegraphics[width=\linewidth]{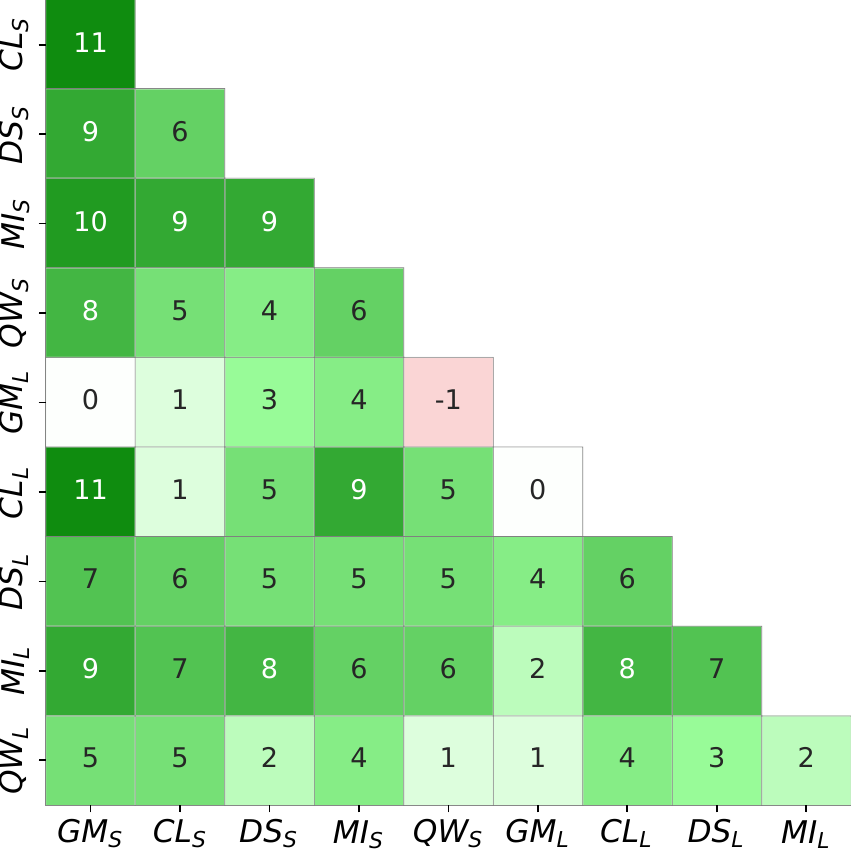}
        \subcaption{HumanEval-Java}\label{fig:heatmap-hej}
    \end{subfigure}
    \begin{subfigure}[b]{0.3\textwidth}
        \includegraphics[width=\linewidth]{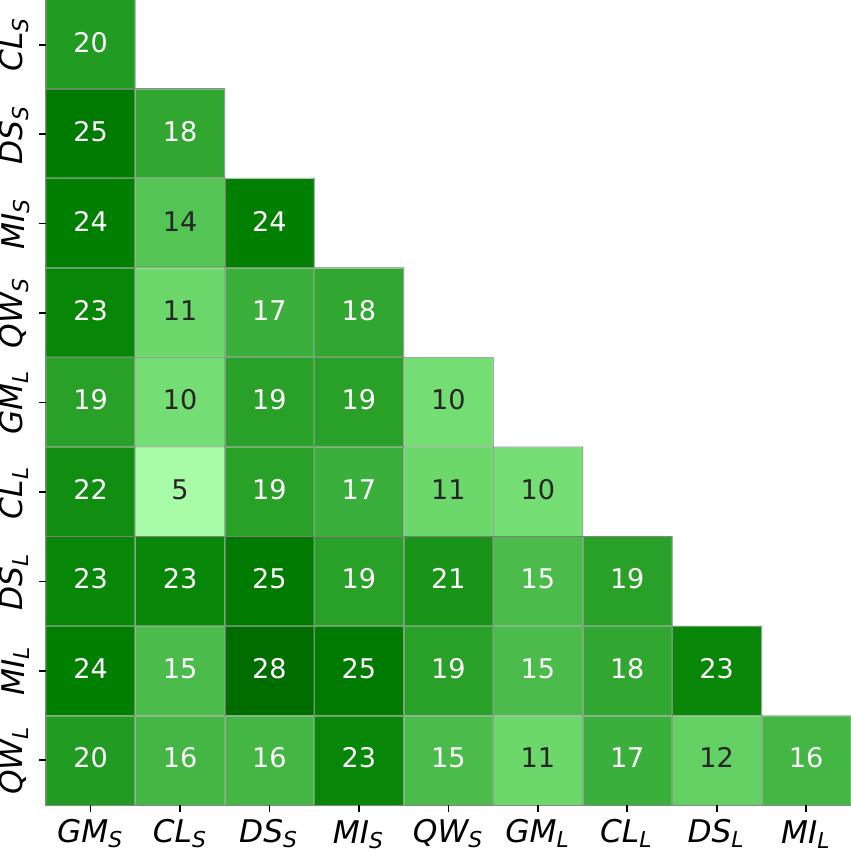}
        \subcaption{Defects4J}\label{fig:heatmap-d4j}
    \end{subfigure}
    \begin{subfigure}[b]{0.3\textwidth}
        \includegraphics[width=\linewidth]{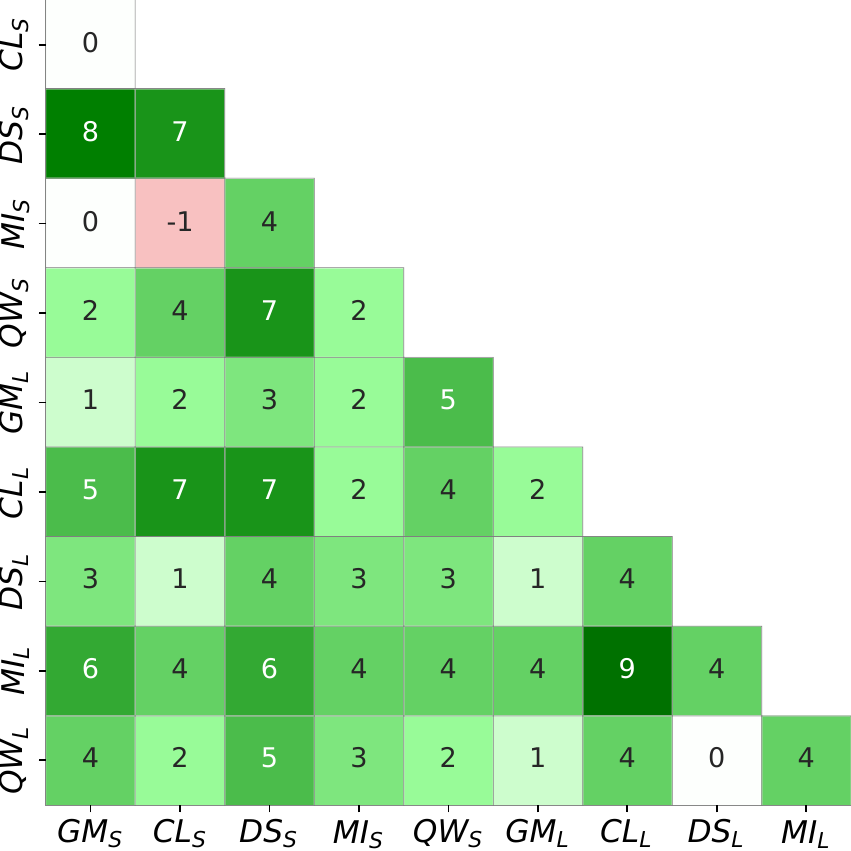}
        \subcaption{LiveCodeBench}\label{fig:heatmap-lcb}
    \end{subfigure}
    \caption{[RQ2.2] Comparison of two-model ensembles with diversity-based strategy using CodeBLEU vs. Naive strategy. The heatmaps show the difference in the number of plausible candidates per benchmark (e.g. a negative score indicates that the CodeBLEU diversity-based strategy performs worse than the Naive ensemble strategy).}
    \label{fig:heatmaps-codebleu-Naive}
    \Description[Short Descr]{Long Descr}
\end{figure*}

\subsubsection{Generalization to smaller ensembles}
While large ensembles show high potential, practical applications may need cost-effective solutions.
Therefore, we study the generalization of the previous strategies to smaller ensembles.
In particular, we analyze ensembles with two models by considering every possible pair of models.
First, we investigate the effectiveness of our Naive baselines, which achieved high performance on ensembles of 5 and 10 models.
\autoref{fig:heatmaps-general-Naive} shows heatmaps indicating the performance difference between the Naive strategy for any ensemble of two models and simply choosing the best single model in the pair.
Negative values indicate cases where the Naive strategy performs worse than the best model of the pair.
On Defects4J, these negative cases are frequent (17 out of 45), reaching a maximum drop of -15 problems when pairing $CL_S$ and $DS_L$.
Similarly, we also note drops of -8 and -6 problems for HumanEval-Java and LiveCodeBench, respectively.
However, across all three benchmarks, we can also notice performance gains by using the Naive strategy.
For instance, the combination of $DS_L$ and $QW_L$ results in an improvement of 22 problems for Defects4J.
Similarly to the negative trend, this positive trend also extends to the other two benchmarks, resulting in an improvement of 12 and 17 problems solved in HumanEval-Java and LiveCodeBench, respectively.
These results demonstrate that any ensemble is not always beneficial, and one needs to be careful when using a Naive ensemble strategy.
Without studying the combinations of the models and without a complex selection strategy, combining models can be actively detrimental to the final outcomes.

\subsubsection{Complex strategies in smaller ensembles}
We apply one of our strategies to these same model pairs.
We have chosen CodeBLEU but the other metrics provide very similar trends.
As a selection strategy, we chose to maximize diversity since it provided the best outcomes on bigger ensembles regardless of the metric chosen (see \autoref{tab:results-all-hej-modified}, \autoref{tab:results-all-d4j-modified}, and \autoref{tab:results-all-lcb-modified}).
\autoref{fig:heatmaps-codebleu-Naive} shows heatmaps indicating the performance difference between our chosen approach and the Naive strategy.
The results clearly indicate that our diversity-driven strategy provides a consistent and substantial performance gain across all model pairs and benchmarks.
Our strategy outperforms or matches the Naive approach for all pairs except two, one case in HumanEval-Java and one case in LiveCodeBench, both cases decreasing the number of problems solved by only one.
Furthermore, for all pairs of models and benchmarks, our strategy provides an improvement over the best score of the two models, 
consistently outperforming the Naive strategy.
The improvements are most pronounced on Defects4J, with over 20 additional problems solved for 16 pairs compared to the Naive strategy.
These results indicate that diversity-based selection is not only effective for larger ensembles, but it is even more effective for smaller ones.
While we present results for only one strategy for brevity, these trends are consistent across other metrics and are available in our public repository.\footref{foot:figshare-link}

\subsubsection{Computational Overhead of Strategies}
The size of the ensemble is one of the key factors indicating the computational cost, however, it is not the only one.
The metrics we have evaluated also add overhead to the calculations.
Output-based metrics require pairwise comparison between all candidate solutions, which scales quadratically with the number of candidates.
As the ensemble size grows, this cost may become a bottleneck.
However, these metrics hold the advantage of being model-agnostic, as they use the generated code directly.
Moreover, their hardware requirements are relatively low since CodeBLEU is CPU-bound and CodeBERTScore can be run effectively on consumer-grade GPUs.

In comparison, confidence-based metrics are more scalable. %
The cost of computing entropy scales linearly with the number of candidates.
This linear efficiency makes them suitable for large ensembles where quadratic scaling may be unfeasible.
However, these metrics depend on the availability of token-level probabilities from the underlying model.
This information is often not disclosed in closed-source models and some commercial APIs, rendering these metrics unfeasible.

\section{Threats to Validity}
One internal validity threat relates to the hyperparameters selection (e.g. in the generation of outputs or for the selection metrics).
We address this by enforcing standard and deterministic generation settings with a temperature of $0$ across all models and default hyperparameters for the selection metrics.
Another threat to internal validity is concerned with potential data leakage from benchmarks into the training data of the models.
This is particularly problematic for older benchmarks such as Defects4J.
We address this threat by including a more recent APR benchmark (i.e. HumanEval-Java) and a code generation benchmark created specifically to avoid data contamination (i.e. LiveCodeBench).

A threat to the external validity of our study is the generalizability of results, for instance, to other models or ensembles, as well as other tasks and programming languages. 
We mitigate this threat by employing a diverse set of ten models from five different families and various parameter sizes.
While exact performance numbers may differ, we believe that our core findings are likely to hold more broadly.
However, confirming this across even bigger and more heterogeneous ensembles remains an important direction for future work.
Similarly, our study uses Java and Python and focuses on code generation and repair, and further work would be needed to confirm our results for other tasks and languages.

The main threat to construct validity is our reliance on plausibility to assess correctness.
Although practical and scalable, this metric does not guarantee true correctness since tests may not fully cover edge cases or outputs may overfit to the tests.
While some prior work in the APR field manually checked all outputs,
this is a labor-intensive and subjective process, as reviewers may decide correctness based on different standards~\cite{wang2021:automated}.
In our experiments, we generate nearly 20,000 plausible candidate solutions which makes manual evaluation unfeasible.
Another threat to construct validity relates to the number of outputs we choose to generate or select from the pool of candidates.
To mitigate this threat, we select this based on observations from related work~\cite{jiang2023:impact,silva2025:repairllama, ruiz2025:art}.

\section{Conclusion}
This work provides empirical evidence that an ensemble of diverse models, if combined effectively,  outperforms individual models on two software engineering tasks.
We systematically explore the complementarity of different LLMs and identify effective heuristics for selecting correct solutions from the aggregated pool of outputs.

Our extensive empirical study employed ten LLMs from five families across three benchmarks for code generation and automatic program repair.
We provide evidence that smaller models can provide valuable solutions that were not found by their larger counterparts.
Moreover, the likelihood of providing diverse solutions is increased if the models belong to different families.
We find that complementarity can improve the ensemble's theoretical potential over its best individual component by as much as 83\% (from 112 to 205 solved problems on Defects4J).
However, we demonstrate that the realization of the potential of the ensemble critically depends on the selection strategy.
A naive selection strategy is unreliable, and heuristics based on consensus often converge on popular but incorrect solutions, a situation we define as the \emph{popularity trap}.
We introduce a diversity-based selection strategy that consistently outperforms consensus, achieving up to 95\% of the ensemble's theoretical potential.
Finally, we extend our analysis to smaller and cost-effective ensembles of two models, where we empirically show that a diversity-based strategy consistently outperforms the results achieved by either of the models and the naive strategy.

Our findings lay the foundation for building more sophisticated and efficient ensembles.
Each component can be selected based on its unique and non-overlapping capabilities.
Furthermore, identifying the unique problem-solving capabilities of each model is key to advancing into more complex multi-agent systems.
For practitioners, our work provides actionable insights: deploying an ensemble of models, especially smaller ones, with a diversity-based selection heuristic results in a consistent performance increase compared to relying on a single model.

\head{Future Work}
There are several ways to extend this work.
First is expanding the scope and scale of our study as indicated in the previous section. 
This includes testing larger and more heterogeneus ensembles, investigating other software engineering tasks (e.g., code summarization, defect detection, or test case generation), and other programming languages.
Another is to investigate alternative metrics and selection strategies. For example, a multi-stage process could use a cheap metric to quickly filter a large pool of candidates and then apply a more expensive metric on the reduced set.

Finally, instead of simply selecting from a static pool of generated outputs, models could be made to interact dynamically in an agentic fashion. 
This could, for example, involve 
(a) iterative refinement~\cite{grishina2025:fully}, where one or more models generate solutions, and other models debug these, creating an iterative refinement cycle, or
(b) task decomposition, where a primary agent breaks down a complex coding problem into sub-tasks that are delegated to specialized models that are best suited for each part.

\section{Data Availability}
The temporary and anonymized replication package for this work is available online.\footnote{\url{https://figshare.com/s/bcb1bfd02cf5ef993e4e}\label{foot:figshare-link}}
This package encompasses: (1) the source code required to replicate the experiments, (2) the generated outputs for all benchmarks and models, (3) the metrics calculated for said outputs, and (4) the scripts to analyze the results and generate figures. 

\begin{acks}
This work is supported by the Research Council of Norway through the secureIT project (IKTPLUSS \#288787), and by the European Union through the Horizon Europe Marie Sk\l{}odowska-Curie Actions (\#101151798).
The empirical evaluation made use of the Experimental Infrastructure for Exploration of Exascale Computing (eX3), 
financially supported by the Research Council of Norway under contract \#270053. 
In addition, we acknowledge Sigma2, Norway for awarding this project access to the LUMI supercomputer, owned by the EuroHPC Joint Undertaking, hosted by CSC (Finland) and the LUMI consortium through the Research Council of Norway. 
\end{acks}

\bibliographystyle{ACM-Reference-Format}
\bibliography{EnsembleModel}


\begin{thebibliography}{46}


\ifx \showCODEN    \undefined \def \showCODEN     #1{\unskip}     \fi
\ifx \showISBNx    \undefined \def \showISBNx     #1{\unskip}     \fi
\ifx \showISBNxiii \undefined \def \showISBNxiii  #1{\unskip}     \fi
\ifx \showISSN     \undefined \def \showISSN      #1{\unskip}     \fi
\ifx \showLCCN     \undefined \def \showLCCN      #1{\unskip}     \fi
\ifx \shownote     \undefined \def \shownote      #1{#1}          \fi
\ifx \showarticletitle \undefined \def \showarticletitle #1{#1}   \fi
\ifx \showURL      \undefined \def \showURL       {\relax}        \fi
\providecommand\bibfield[2]{#2}
\providecommand\bibinfo[2]{#2}
\providecommand\natexlab[1]{#1}
\providecommand\showeprint[2][]{arXiv:#2}

\bibitem[Ashiga et~al\mbox{.}(2025)]%
        {ashiga2025:ensemble}
\bibfield{author}{\bibinfo{person}{Mari Ashiga}, \bibinfo{person}{Wei Jie},
  \bibinfo{person}{Fan Wu}, \bibinfo{person}{Vardan Voskanyan},
  \bibinfo{person}{Fateme Dinmohammadi}, \bibinfo{person}{Paul Brookes},
  \bibinfo{person}{Jingzhi Gong}, {and} \bibinfo{person}{Zheng Wang}.}
  \bibinfo{year}{2025}\natexlab{}.
\newblock \bibinfo{title}{Ensemble Learning for Large Language Models in Text
  and Code Generation: A Survey}.
\newblock
\showeprint[arxiv]{2503.13505}~[cs]
\href{https://doi.org/10.48550/arXiv.2503.13505}{doi:\nolinkurl{10.48550/arXiv.2503.13505}}


\bibitem[Chen et~al\mbox{.}(2025)]%
        {chen2025:debatecoder}
\bibfield{author}{\bibinfo{person}{Jizheng Chen}, \bibinfo{person}{Kounianhua
  Du}, \bibinfo{person}{Xinyi Dai}, \bibinfo{person}{Weiming Zhang},
  \bibinfo{person}{Xihuai Wang}, \bibinfo{person}{Yasheng Wang},
  \bibinfo{person}{Ruiming Tang}, \bibinfo{person}{Weinan Zhang}, {and}
  \bibinfo{person}{Yong Yu}.} \bibinfo{year}{2025}\natexlab{}.
\newblock \showarticletitle{{{DebateCoder}}: {{Towards Collective
  Intelligence}} of {{LLMs}} via {{Test Case Driven LLM Debate}} for {{Code
  Generation}}}. In \bibinfo{booktitle}{\emph{Proceedings of the 63rd {{Annual
  Meeting}} of the {{Association}} for {{Computational Linguistics}}
  ({{Volume}} 1: {{Long Papers}})}},
  \bibfield{editor}{\bibinfo{person}{Wanxiang Che}, \bibinfo{person}{Joyce
  Nabende}, \bibinfo{person}{Ekaterina Shutova}, {and}
  \bibinfo{person}{Mohammad~Taher Pilehvar}} (Eds.).
  \bibinfo{publisher}{Association for Computational Linguistics},
  \bibinfo{address}{Vienna, Austria}, \bibinfo{pages}{12055--12065}.
\newblock
\showISBNx{979-8-89176-251-0}
\href{https://doi.org/10.18653/v1/2025.acl-long.589}{doi:\nolinkurl{10.18653/v1/2025.acl-long.589}}


\bibitem[Dou et~al\mbox{.}(2024)]%
        {dou2024:whats}
\bibfield{author}{\bibinfo{person}{Shihan Dou}, \bibinfo{person}{Haoxiang Jia},
  \bibinfo{person}{Shenxi Wu}, \bibinfo{person}{Huiyuan Zheng},
  \bibinfo{person}{Weikang Zhou}, \bibinfo{person}{Muling Wu},
  \bibinfo{person}{Mingxu Chai}, \bibinfo{person}{Jessica Fan},
  \bibinfo{person}{Caishuang Huang}, \bibinfo{person}{Yunbo Tao},
  \bibinfo{person}{Yan Liu}, \bibinfo{person}{Enyu Zhou}, \bibinfo{person}{Ming
  Zhang}, \bibinfo{person}{Yuhao Zhou}, \bibinfo{person}{Yueming Wu},
  \bibinfo{person}{Rui Zheng}, \bibinfo{person}{Ming Wen},
  \bibinfo{person}{Rongxiang Weng}, \bibinfo{person}{Jingang Wang},
  \bibinfo{person}{Xunliang Cai}, \bibinfo{person}{Tao Gui},
  \bibinfo{person}{Xipeng Qiu}, \bibinfo{person}{Qi Zhang}, {and}
  \bibinfo{person}{Xuanjing Huang}.} \bibinfo{year}{2024}\natexlab{}.
\newblock \bibinfo{title}{What's Wrong with Your Code Generated by Large
  Language Models? {{An}} Extensive Study}.
\newblock
\showeprint[arxiv]{2407.06153}~[cs]
\href{https://doi.org/10.48550/arXiv.2407.06153}{doi:\nolinkurl{10.48550/arXiv.2407.06153}}


\bibitem[Du et~al\mbox{.}(2021)]%
        {du2021:single}
\bibfield{author}{\bibinfo{person}{Lun Du}, \bibinfo{person}{Xiaozhou Shi},
  \bibinfo{person}{Yanlin Wang}, \bibinfo{person}{Ensheng Shi},
  \bibinfo{person}{Shi Han}, {and} \bibinfo{person}{Dongmei Zhang}.}
  \bibinfo{year}{2021}\natexlab{}.
\newblock \showarticletitle{Is a {{Single Model Enough}}? {{MuCoS}}: {{A
  Multi-Model Ensemble Learning Approach}} for {{Semantic Code Search}}}. In
  \bibinfo{booktitle}{\emph{Proceedings of the 30th {{ACM International
  Conference}} on {{Information}} \& {{Knowledge Management}}}}.
  \bibinfo{publisher}{ACM}, \bibinfo{address}{Virtual Event Queensland
  Australia}, \bibinfo{pages}{2994--2998}.
\newblock
\showISBNx{978-1-4503-8446-9}
\href{https://doi.org/10.1145/3459637.3482127}{doi:\nolinkurl{10.1145/3459637.3482127}}


\bibitem[Du et~al\mbox{.}(2024)]%
        {du2024:evaluating}
\bibfield{author}{\bibinfo{person}{Xueying Du}, \bibinfo{person}{Mingwei Liu},
  \bibinfo{person}{Kaixin Wang}, \bibinfo{person}{Hanlin Wang},
  \bibinfo{person}{Junwei Liu}, \bibinfo{person}{Yixuan Chen},
  \bibinfo{person}{Jiayi Feng}, \bibinfo{person}{Chaofeng Sha},
  \bibinfo{person}{Xin Peng}, {and} \bibinfo{person}{Yiling Lou}.}
  \bibinfo{year}{2024}\natexlab{}.
\newblock \showarticletitle{Evaluating Large Language Models in Class-Level
  Code Generation}. In \bibinfo{booktitle}{\emph{Proceedings of the
  {{IEEE}}/{{ACM}} 46th {{International Conference}} on {{Software
  Engineering}}}} \emph{(\bibinfo{series}{{{ICSE}} '24})}.
  \bibinfo{publisher}{Association for Computing Machinery},
  \bibinfo{address}{New York, NY, USA}, \bibinfo{pages}{1--13}.
\newblock
\showISBNx{979-8-4007-0217-4}
\href{https://doi.org/10.1145/3597503.3639219}{doi:\nolinkurl{10.1145/3597503.3639219}}


\bibitem[Fan et~al\mbox{.}(2023)]%
        {fan2023:large}
\bibfield{author}{\bibinfo{person}{Angela Fan}, \bibinfo{person}{Beliz
  Gokkaya}, \bibinfo{person}{Mark Harman}, \bibinfo{person}{Mitya Lyubarskiy},
  \bibinfo{person}{Shubho Sengupta}, \bibinfo{person}{Shin Yoo}, {and}
  \bibinfo{person}{Jie~M. Zhang}.} \bibinfo{year}{2023}\natexlab{}.
\newblock \showarticletitle{Large Language Models for Software Engineering:
  Survey and Open Problems}. In \bibinfo{booktitle}{\emph{2023 {{IEEE}}/{{ACM
  International Conference}} on {{Software Engineering}}: {{Future}} of
  {{Software Engineering}} (Icse-Fose)}}. \bibinfo{pages}{31--53}.
\newblock
\href{https://doi.org/10.1109/ICSE-FoSE59343.2023.00008}{doi:\nolinkurl{10.1109/ICSE-FoSE59343.2023.00008}}


\bibitem[Feng et~al\mbox{.}(2020)]%
        {feng2020:codebert}
\bibfield{author}{\bibinfo{person}{Zhangyin Feng}, \bibinfo{person}{Daya Guo},
  \bibinfo{person}{Duyu Tang}, \bibinfo{person}{Nan Duan},
  \bibinfo{person}{Xiaocheng Feng}, \bibinfo{person}{Ming Gong},
  \bibinfo{person}{Linjun Shou}, \bibinfo{person}{Bing Qin},
  \bibinfo{person}{Ting Liu}, \bibinfo{person}{Daxin Jiang}, {and}
  \bibinfo{person}{Ming Zhou}.} \bibinfo{year}{2020}\natexlab{}.
\newblock \showarticletitle{{{CodeBERT}}: A Pre-Trained Model for Programming
  and Natural Languages}. In \bibinfo{booktitle}{\emph{Findings of the
  {{Association}} for {{Computational Linguistics}}: {{EMNLP}} 2020}},
  \bibfield{editor}{\bibinfo{person}{Trevor Cohn}, \bibinfo{person}{Yulan He},
  {and} \bibinfo{person}{Yang Liu}} (Eds.). \bibinfo{publisher}{Association for
  Computational Linguistics}, \bibinfo{address}{Online},
  \bibinfo{pages}{1536--1547}.
\newblock
\href{https://doi.org/10.18653/v1/2020.findings-emnlp.139}{doi:\nolinkurl{10.18653/v1/2020.findings-emnlp.139}}


\bibitem[Grishina et~al\mbox{.}(2025)]%
        {grishina2025:fully}
\bibfield{author}{\bibinfo{person}{Anastasiia Grishina}, \bibinfo{person}{Vadim
  Liventsev}, \bibinfo{person}{Aki H{\"a}rm{\"a}}, {and} \bibinfo{person}{Leon
  Moonen}.} \bibinfo{year}{2025}\natexlab{}.
\newblock \showarticletitle{Fully Autonomous Programming Using Iterative
  Multi-Agent Debugging with Large Language Models}.
\newblock \bibinfo{journal}{\emph{ACM Trans. Evol. Learn. Optim.}}
  \bibinfo{volume}{5}, \bibinfo{number}{1} (\bibinfo{date}{March}
  \bibinfo{year}{2025}), \bibinfo{pages}{8:1--8:37}.
\newblock
\href{https://doi.org/10.1145/3719351}{doi:\nolinkurl{10.1145/3719351}}


\bibitem[Hassid et~al\mbox{.}(2024)]%
        {hassid2024:larger}
\bibfield{author}{\bibinfo{person}{Michael Hassid}, \bibinfo{person}{Tal
  Remez}, \bibinfo{person}{Jonas Gehring}, \bibinfo{person}{Roy Schwartz},
  {and} \bibinfo{person}{Yossi Adi}.} \bibinfo{year}{2024}\natexlab{}.
\newblock \bibinfo{title}{The Larger the Better? {{Improved LLM}}
  Code-Generation via Budget Reallocation}.
\newblock
\showeprint[arxiv]{2404.00725}~[cs]
\href{https://doi.org/10.48550/arXiv.2404.00725}{doi:\nolinkurl{10.48550/arXiv.2404.00725}}


\bibitem[He et~al\mbox{.}(2025)]%
        {he2025:llmbased}
\bibfield{author}{\bibinfo{person}{Junda He}, \bibinfo{person}{Christoph
  Treude}, {and} \bibinfo{person}{David Lo}.} \bibinfo{year}{2025}\natexlab{}.
\newblock \showarticletitle{{{LLM-based}} Multi-Agent Systems for Software
  Engineering: Literature Review, Vision, and the Road Ahead}.
\newblock \bibinfo{journal}{\emph{ACM Trans. Softw. Eng. Methodol.}}
  \bibinfo{volume}{34}, \bibinfo{number}{5} (\bibinfo{date}{May}
  \bibinfo{year}{2025}), \bibinfo{pages}{124:1--124:30}.
\newblock
\showISSN{1049-331X}
\href{https://doi.org/10.1145/3712003}{doi:\nolinkurl{10.1145/3712003}}


\bibitem[Hort et~al\mbox{.}(2025)]%
        {hort2025:semanticpreserving}
\bibfield{author}{\bibinfo{person}{Max Hort}, \bibinfo{person}{Linas
  Vidziunas}, {and} \bibinfo{person}{Leon Moonen}.}
  \bibinfo{year}{2025}\natexlab{}.
\newblock \showarticletitle{Semantic-Preserving Transformations as Mutation
  Operators: A Study on Their Effectiveness in Defect Detection}. In
  \bibinfo{booktitle}{\emph{2025 {{IEEE International Conference}} on
  {{Software Testing}}, {{Verification}} and {{Validation Workshops}}
  ({{ICSTW}})}}. \bibinfo{publisher}{IEEE Computer Society},
  \bibinfo{pages}{337--346}.
\newblock
\showISBNx{979-8-3315-3467-7}
\showISSN{2159-4848}
\href{https://doi.org/10.1109/ICSTW64639.2025.10962512}{doi:\nolinkurl{10.1109/ICSTW64639.2025.10962512}}


\bibitem[Hou et~al\mbox{.}(2024)]%
        {hou2024:large}
\bibfield{author}{\bibinfo{person}{Xinyi Hou}, \bibinfo{person}{Yanjie Zhao},
  \bibinfo{person}{Yue Liu}, \bibinfo{person}{Zhou Yang},
  \bibinfo{person}{Kailong Wang}, \bibinfo{person}{Li Li},
  \bibinfo{person}{Xiapu Luo}, \bibinfo{person}{David Lo},
  \bibinfo{person}{John Grundy}, {and} \bibinfo{person}{Haoyu Wang}.}
  \bibinfo{year}{2024}\natexlab{}.
\newblock \showarticletitle{Large Language Models for Software Engineering: A
  Systematic Literature Review}.
\newblock \bibinfo{journal}{\emph{ACM Trans. Softw. Eng. Methodol.}}
  \bibinfo{volume}{33}, \bibinfo{number}{8} (\bibinfo{date}{Dec.}
  \bibinfo{year}{2024}), \bibinfo{pages}{220:1--220:79}.
\newblock
\showISSN{1049-331X}
\href{https://doi.org/10.1145/3695988}{doi:\nolinkurl{10.1145/3695988}}


\bibitem[Jain et~al\mbox{.}(2024)]%
        {jain2024:livecodebench}
\bibfield{author}{\bibinfo{person}{Naman Jain}, \bibinfo{person}{King Han},
  \bibinfo{person}{Alex Gu}, \bibinfo{person}{Wen-Ding Li},
  \bibinfo{person}{Fanjia Yan}, \bibinfo{person}{Tianjun Zhang},
  \bibinfo{person}{Sida Wang}, \bibinfo{person}{Armando {Solar-Lezama}},
  \bibinfo{person}{Koushik Sen}, {and} \bibinfo{person}{Ion Stoica}.}
  \bibinfo{year}{2024}\natexlab{}.
\newblock \showarticletitle{{{LiveCodeBench}}: Holistic and Contamination Free
  Evaluation of Large Language Models for Code}. In
  \bibinfo{booktitle}{\emph{The {{Thirteenth International Conference}} on
  {{Learning Representations}}}}.
\newblock
\urldef\tempurl%
\url{https://openreview.net/forum?id=chfJJYC3iL}
\showURL{%
\tempurl}


\bibitem[Jiang et~al\mbox{.}(2023b)]%
        {jiang2023:llmblender}
\bibfield{author}{\bibinfo{person}{Dongfu Jiang}, \bibinfo{person}{Xiang Ren},
  {and} \bibinfo{person}{Bill~Yuchen Lin}.} \bibinfo{year}{2023}\natexlab{b}.
\newblock \bibinfo{title}{{{LLM-blender}}: Ensembling Large Language Models
  with Pairwise Ranking and Generative Fusion}.
\newblock
\showeprint[arxiv]{2306.02561}~[cs]
\href{https://doi.org/10.48550/arXiv.2306.02561}{doi:\nolinkurl{10.48550/arXiv.2306.02561}}


\bibitem[Jiang et~al\mbox{.}(2023a)]%
        {jiang2023:impact}
\bibfield{author}{\bibinfo{person}{Nan Jiang}, \bibinfo{person}{Kevin Liu},
  \bibinfo{person}{Thibaud Lutellier}, {and} \bibinfo{person}{Lin Tan}.}
  \bibinfo{year}{2023}\natexlab{a}.
\newblock \showarticletitle{Impact of Code Language Models on Automated Program
  Repair}. In \bibinfo{booktitle}{\emph{Proceedings of the 45th {{International
  Conference}} on {{Software Engineering}}}} \emph{(\bibinfo{series}{{{ICSE}}
  '23})}. \bibinfo{publisher}{IEEE Press}, \bibinfo{address}{Melbourne,
  Victoria, Australia}, \bibinfo{pages}{1430--1442}.
\newblock
\showISBNx{978-1-6654-5701-9}
\href{https://doi.org/10.1109/ICSE48619.2023.00125}{doi:\nolinkurl{10.1109/ICSE48619.2023.00125}}


\bibitem[Just et~al\mbox{.}(2014)]%
        {just2014:defects4j}
\bibfield{author}{\bibinfo{person}{Ren{\'e} Just}, \bibinfo{person}{Darioush
  Jalali}, {and} \bibinfo{person}{Michael~D. Ernst}.}
  \bibinfo{year}{2014}\natexlab{}.
\newblock \showarticletitle{{{Defects4J}}: A Database of Existing Faults to
  Enable Controlled Testing Studies for Java Programs}. In
  \bibinfo{booktitle}{\emph{23th {{International Symposium}} on {{Software
  Testing}} and {{Analysis}} ({{ISSTA}})}} \emph{(\bibinfo{series}{{{ISSTA}}
  2014})}. \bibinfo{publisher}{ACM}, \bibinfo{pages}{437--440}.
\newblock
\showISBNx{978-1-4503-2645-2}
\href{https://doi.org/10.1145/2610384.2628055}{doi:\nolinkurl{10.1145/2610384.2628055}}


\bibitem[Kaplan et~al\mbox{.}(2020)]%
        {kaplan2020:scaling}
\bibfield{author}{\bibinfo{person}{Jared Kaplan}, \bibinfo{person}{Sam
  McCandlish}, \bibinfo{person}{Tom Henighan}, \bibinfo{person}{Tom~B. Brown},
  \bibinfo{person}{Benjamin Chess}, \bibinfo{person}{Rewon Child},
  \bibinfo{person}{Scott Gray}, \bibinfo{person}{Alec Radford},
  \bibinfo{person}{Jeffrey Wu}, {and} \bibinfo{person}{Dario Amodei}.}
  \bibinfo{year}{2020}\natexlab{}.
\newblock \bibinfo{title}{Scaling Laws for Neural Language Models}.
\newblock
\showeprint[arxiv]{2001.08361}~[cs]
\href{https://doi.org/10.48550/arXiv.2001.08361}{doi:\nolinkurl{10.48550/arXiv.2001.08361}}


\bibitem[Liu et~al\mbox{.}(2020)]%
        {liu2020:efficiency}
\bibfield{author}{\bibinfo{person}{Kui Liu}, \bibinfo{person}{Shangwen Wang},
  \bibinfo{person}{Anil Koyuncu}, \bibinfo{person}{Kisub Kim},
  \bibinfo{person}{Tegawend{\'e}~F. Bissyand{\'e}}, \bibinfo{person}{Dongsun
  Kim}, \bibinfo{person}{Peng Wu}, \bibinfo{person}{Jacques Klein},
  \bibinfo{person}{Xiaoguang Mao}, {and} \bibinfo{person}{Yves~Le Traon}.}
  \bibinfo{year}{2020}\natexlab{}.
\newblock \showarticletitle{On the Efficiency of Test Suite Based Program
  Repair: A Systematic Assessment of 16 Automated Repair Systems for Java
  Programs}. In \bibinfo{booktitle}{\emph{Proceedings of the {{ACM}}/{{IEEE}}
  42nd {{International Conference}} on {{Software Engineering}}}}
  \emph{(\bibinfo{series}{{{ICSE}} '20})}. \bibinfo{publisher}{Association for
  Computing Machinery}, \bibinfo{address}{New York, NY, USA},
  \bibinfo{pages}{615--627}.
\newblock
\showISBNx{978-1-4503-7121-6}
\href{https://doi.org/10.1145/3377811.3380338}{doi:\nolinkurl{10.1145/3377811.3380338}}


\bibitem[Liu and Liu(2021)]%
        {liu2021:simcls}
\bibfield{author}{\bibinfo{person}{Yixin Liu} {and} \bibinfo{person}{Pengfei
  Liu}.} \bibinfo{year}{2021}\natexlab{}.
\newblock \bibinfo{title}{{{SimCLS}}: {{A Simple Framework}} for {{Contrastive
  Learning}} of {{Abstractive Summarization}}}.
\newblock
\showeprint[arxiv]{2106.01890}~[cs]
\href{https://doi.org/10.48550/arXiv.2106.01890}{doi:\nolinkurl{10.48550/arXiv.2106.01890}}


\bibitem[Lu et~al\mbox{.}(2024)]%
        {lu2024:merge}
\bibfield{author}{\bibinfo{person}{Jinliang Lu}, \bibinfo{person}{Ziliang
  Pang}, \bibinfo{person}{Min Xiao}, \bibinfo{person}{Yaochen Zhu},
  \bibinfo{person}{Rui Xia}, {and} \bibinfo{person}{Jiajun Zhang}.}
  \bibinfo{year}{2024}\natexlab{}.
\newblock \bibinfo{title}{Merge, Ensemble, and Cooperate! {{A}} Survey on
  Collaborative Strategies in the Era of Large Language Models}.
\newblock
\showeprint[arxiv]{2407.06089}
\href{https://doi.org/10.48550/ARXIV.2407.06089}{doi:\nolinkurl{10.48550/ARXIV.2407.06089}}


\bibitem[Masoudnia and Ebrahimpour(2014)]%
        {masoudnia2014:mixture}
\bibfield{author}{\bibinfo{person}{Saeed Masoudnia} {and} \bibinfo{person}{Reza
  Ebrahimpour}.} \bibinfo{year}{2014}\natexlab{}.
\newblock \showarticletitle{Mixture of Experts: A Literature Survey}.
\newblock \bibinfo{journal}{\emph{Artificial Intelligence Review}}
  \bibinfo{volume}{42}, \bibinfo{number}{2} (\bibinfo{date}{Aug.}
  \bibinfo{year}{2014}), \bibinfo{pages}{275--293}.
\newblock
\showISSN{0269-2821, 1573-7462}
\href{https://doi.org/10.1007/s10462-012-9338-y}{doi:\nolinkurl{10.1007/s10462-012-9338-y}}


\bibitem[Mienye and Sun(2022)]%
        {mienye2022:survey}
\bibfield{author}{\bibinfo{person}{Ibomoiye~Domor Mienye} {and}
  \bibinfo{person}{Yanxia Sun}.} \bibinfo{year}{2022}\natexlab{}.
\newblock \showarticletitle{A Survey of Ensemble Learning: Concepts,
  Algorithms, Applications, and Prospects}.
\newblock \bibinfo{journal}{\emph{IEEE Access}}  \bibinfo{volume}{10}
  (\bibinfo{year}{2022}), \bibinfo{pages}{99129--99149}.
\newblock
\showISSN{2169-3536}
\href{https://doi.org/10.1109/ACCESS.2022.3207287}{doi:\nolinkurl{10.1109/ACCESS.2022.3207287}}


\bibitem[Noller et~al\mbox{.}(2022)]%
        {noller2022:trust}
\bibfield{author}{\bibinfo{person}{Yannic Noller}, \bibinfo{person}{Ridwan
  Shariffdeen}, \bibinfo{person}{Xiang Gao}, {and} \bibinfo{person}{Abhik
  Roychoudhury}.} \bibinfo{year}{2022}\natexlab{}.
\newblock \showarticletitle{Trust Enhancement Issues in Program Repair}. In
  \bibinfo{booktitle}{\emph{Proceedings of the 44th {{International
  Conference}} on {{Software Engineering}}}} \emph{(\bibinfo{series}{{{ICSE}}
  '22})}. \bibinfo{publisher}{Association for Computing Machinery},
  \bibinfo{address}{New York, NY, USA}, \bibinfo{pages}{2228--2240}.
\newblock
\showISBNx{978-1-4503-9221-1}
\href{https://doi.org/10.1145/3510003.3510040}{doi:\nolinkurl{10.1145/3510003.3510040}}


\bibitem[Papineni et~al\mbox{.}(2002)]%
        {papineni2002:bleu}
\bibfield{author}{\bibinfo{person}{Kishore Papineni}, \bibinfo{person}{Salim
  Roukos}, \bibinfo{person}{Todd Ward}, {and} \bibinfo{person}{Wei-Jing Zhu}.}
  \bibinfo{year}{2002}\natexlab{}.
\newblock \showarticletitle{Bleu: A Method for Automatic Evaluation of Machine
  Translation}. In \bibinfo{booktitle}{\emph{Proceedings of the 40th {{Annual
  Meeting}} of the {{Association}} for {{Computational Linguistics}}}},
  \bibfield{editor}{\bibinfo{person}{Pierre Isabelle}, \bibinfo{person}{Eugene
  Charniak}, {and} \bibinfo{person}{Dekang Lin}} (Eds.).
  \bibinfo{publisher}{Association for Computational Linguistics},
  \bibinfo{address}{Philadelphia, Pennsylvania, USA},
  \bibinfo{pages}{311--318}.
\newblock
\href{https://doi.org/10.3115/1073083.1073135}{doi:\nolinkurl{10.3115/1073083.1073135}}


\bibitem[Rasheed et~al\mbox{.}(2025)]%
        {rasheed2025:large}
\bibfield{author}{\bibinfo{person}{Zeeshan Rasheed}, \bibinfo{person}{Muhammad
  Waseem}, \bibinfo{person}{Kai~Kristian Kemell}, \bibinfo{person}{Aakash
  Ahmad}, \bibinfo{person}{Malik~Abdul Sami}, \bibinfo{person}{Jussi Rasku},
  \bibinfo{person}{Kari Syst{\"a}}, {and} \bibinfo{person}{Pekka Abrahamsson}.}
  \bibinfo{year}{2025}\natexlab{}.
\newblock \bibinfo{title}{Large Language Models for Code Generation: The
  Practitioners Perspective}.
\newblock
\showeprint[arxiv]{2501.16998}~[cs]
\href{https://doi.org/10.48550/arXiv.2501.16998}{doi:\nolinkurl{10.48550/arXiv.2501.16998}}


\bibitem[Ravaut et~al\mbox{.}(2023)]%
        {ravaut2023:summareranker}
\bibfield{author}{\bibinfo{person}{Mathieu Ravaut}, \bibinfo{person}{Shafiq
  Joty}, {and} \bibinfo{person}{Nancy~F. Chen}.}
  \bibinfo{year}{2023}\natexlab{}.
\newblock \bibinfo{title}{{{SummaReranker}}: {{A Multi-Task Mixture-of-Experts
  Re-ranking Framework}} for {{Abstractive Summarization}}}.
\newblock
\showeprint[arxiv]{2203.06569}~[cs]
\href{https://doi.org/10.48550/arXiv.2203.06569}{doi:\nolinkurl{10.48550/arXiv.2203.06569}}


\bibitem[Ren et~al\mbox{.}(2020)]%
        {ren2020:codebleu}
\bibfield{author}{\bibinfo{person}{Shuo Ren}, \bibinfo{person}{Daya Guo},
  \bibinfo{person}{Shuai Lu}, \bibinfo{person}{Long Zhou},
  \bibinfo{person}{Shujie Liu}, \bibinfo{person}{Duyu Tang},
  \bibinfo{person}{Neel Sundaresan}, \bibinfo{person}{Ming Zhou},
  \bibinfo{person}{Ambrosio Blanco}, {and} \bibinfo{person}{Shuai Ma}.}
  \bibinfo{year}{2020}\natexlab{}.
\newblock \bibinfo{title}{{{CodeBLEU}}: A Method for Automatic Evaluation of
  Code Synthesis}.
\newblock
\showeprint[arxiv]{2009.10297}~[cs]
\href{https://doi.org/10.48550/arXiv.2009.10297}{doi:\nolinkurl{10.48550/arXiv.2009.10297}}


\bibitem[Ridoy et~al\mbox{.}(2024)]%
        {ridoy2024:enstack}
\bibfield{author}{\bibinfo{person}{Shahriyar~Zaman Ridoy},
  \bibinfo{person}{{\relax Md}. Shazzad Hossain~Shaon},
  \bibinfo{person}{Alfredo Cuzzocrea}, {and} \bibinfo{person}{Mst~Shapna
  Akter}.} \bibinfo{year}{2024}\natexlab{}.
\newblock \showarticletitle{{{EnStack}}: {{An Ensemble Stacking Framework}} of
  {{Large Language Models}} for {{Enhanced Vulnerability Detection}} in
  {{Source Code}}}. In \bibinfo{booktitle}{\emph{2024 {{IEEE International
  Conference}} on {{Big Data}} ({{BigData}})}}. \bibinfo{pages}{6356--6364}.
\newblock
\showISSN{2573-2978}
\href{https://doi.org/10.1109/BigData62323.2024.10825609}{doi:\nolinkurl{10.1109/BigData62323.2024.10825609}}


\bibitem[Ruiz et~al\mbox{.}(2025)]%
        {ruiz2025:art}
\bibfield{author}{\bibinfo{person}{Fernando~Vallecillos Ruiz},
  \bibinfo{person}{Max Hort}, {and} \bibinfo{person}{Leon Moonen}.}
  \bibinfo{year}{2025}\natexlab{}.
\newblock \bibinfo{title}{The Art of Repair: Optimizing Iterative Program
  Repair with Instruction-Tuned Models}.
\newblock
\showeprint[arxiv]{2505.02931}~[cs]
\href{https://doi.org/10.48550/arXiv.2505.02931}{doi:\nolinkurl{10.48550/arXiv.2505.02931}}


\bibitem[Salazar et~al\mbox{.}(2020)]%
        {salazar2020:masked}
\bibfield{author}{\bibinfo{person}{Julian Salazar}, \bibinfo{person}{Davis
  Liang}, \bibinfo{person}{Toan~Q. Nguyen}, {and} \bibinfo{person}{Katrin
  Kirchhoff}.} \bibinfo{year}{2020}\natexlab{}.
\newblock \showarticletitle{Masked {{Language Model Scoring}}}. In
  \bibinfo{booktitle}{\emph{Proceedings of the 58th {{Annual Meeting}} of the
  {{Association}} for {{Computational Linguistics}}}}.
  \bibinfo{pages}{2699--2712}.
\newblock
\showeprint[arxiv]{1910.14659}~[cs]
\href{https://doi.org/10.18653/v1/2020.acl-main.240}{doi:\nolinkurl{10.18653/v1/2020.acl-main.240}}


\bibitem[Sellam et~al\mbox{.}(2020)]%
        {sellam2020:bleurt}
\bibfield{author}{\bibinfo{person}{Thibault Sellam}, \bibinfo{person}{Dipanjan
  Das}, {and} \bibinfo{person}{Ankur~P. Parikh}.}
  \bibinfo{year}{2020}\natexlab{}.
\newblock \bibinfo{title}{{{BLEURT}}: {{Learning Robust Metrics}} for {{Text
  Generation}}}.
\newblock
\showeprint[arxiv]{2004.04696}~[cs]
\href{https://doi.org/10.48550/arXiv.2004.04696}{doi:\nolinkurl{10.48550/arXiv.2004.04696}}


\bibitem[Shypula et~al\mbox{.}(2025)]%
        {shypula2025:evaluating}
\bibfield{author}{\bibinfo{person}{Alexander Shypula}, \bibinfo{person}{Shuo
  Li}, \bibinfo{person}{Botong Zhang}, \bibinfo{person}{Vishakh Padmakumar},
  \bibinfo{person}{Kayo Yin}, {and} \bibinfo{person}{Osbert Bastani}.}
  \bibinfo{year}{2025}\natexlab{}.
\newblock \showarticletitle{Evaluating the Diversity and Quality of {{LLM}}
  Generated Content}. In \bibinfo{booktitle}{\emph{Second {{Conference}} on
  {{Language Modeling}}}}.
\newblock
\urldef\tempurl%
\url{https://openreview.net/forum?id=O7bF6nlSOD#discussion}
\showURL{%
\tempurl}


\bibitem[Silva et~al\mbox{.}(2025)]%
        {silva2025:repairllama}
\bibfield{author}{\bibinfo{person}{Andr{\'e} Silva}, \bibinfo{person}{Sen
  Fang}, {and} \bibinfo{person}{Martin Monperrus}.}
  \bibinfo{year}{2025}\natexlab{}.
\newblock \showarticletitle{{{RepairLLaMA}}: Efficient Representations and
  Fine-Tuned Adapters for Program Repair}.
\newblock \bibinfo{journal}{\emph{IEEE Transactions on Software Engineering}}
  \bibinfo{volume}{51}, \bibinfo{number}{8} (\bibinfo{date}{Aug.}
  \bibinfo{year}{2025}), \bibinfo{pages}{2366--2380}.
\newblock
\showISSN{1939-3520}
\href{https://doi.org/10.1109/TSE.2025.3581062}{doi:\nolinkurl{10.1109/TSE.2025.3581062}}


\bibitem[Sober(2015)]%
        {sober2015:ockhams}
\bibfield{author}{\bibinfo{person}{Elliott Sober}.}
  \bibinfo{year}{2015}\natexlab{}.
\newblock \bibinfo{booktitle}{\emph{Ockham's Razors: A User's Manual}}.
\newblock \bibinfo{publisher}{Cambridge University Press},
  \bibinfo{address}{Cambridge}.
\newblock
\showISBNx{978-1-107-06849-0}
\href{https://doi.org/10.1017/CBO9781107705937}{doi:\nolinkurl{10.1017/CBO9781107705937}}


\bibitem[Sun et~al\mbox{.}(2025)]%
        {sun2025:ensembling}
\bibfield{author}{\bibinfo{person}{Zhihong Sun}, \bibinfo{person}{Jia Li},
  \bibinfo{person}{Yao Wan}, \bibinfo{person}{Chuanyi Li},
  \bibinfo{person}{Hongyu Zhang}, \bibinfo{person}{Zhi Jin},
  \bibinfo{person}{Ge Li}, \bibinfo{person}{Hong Liu}, \bibinfo{person}{Chen
  Lyu}, {and} \bibinfo{person}{Songlin Hu}.} \bibinfo{year}{2025}\natexlab{}.
\newblock \bibinfo{title}{Ensembling Large Language Models for Code
  Vulnerability Detection: An Empirical Evaluation}.
\newblock
\showeprint[arxiv]{2509.12629}~[cs]
\href{https://doi.org/10.48550/arXiv.2509.12629}{doi:\nolinkurl{10.48550/arXiv.2509.12629}}


\bibitem[Wang et~al\mbox{.}(2021)]%
        {wang2021:automated}
\bibfield{author}{\bibinfo{person}{Shangwen Wang}, \bibinfo{person}{Ming Wen},
  \bibinfo{person}{Bo Lin}, \bibinfo{person}{Hongjun Wu},
  \bibinfo{person}{Yihao Qin}, \bibinfo{person}{Deqing Zou},
  \bibinfo{person}{Xiaoguang Mao}, {and} \bibinfo{person}{Hai Jin}.}
  \bibinfo{year}{2021}\natexlab{}.
\newblock \showarticletitle{Automated Patch Correctness Assessment: How Far Are
  We?}. In \bibinfo{booktitle}{\emph{Proceedings of the 35th {{IEEE}}/{{ACM
  International Conference}} on {{Automated Software Engineering}}}}
  \emph{(\bibinfo{series}{{{ASE}} '20})}. \bibinfo{publisher}{Association for
  Computing Machinery}, \bibinfo{address}{New York, NY, USA},
  \bibinfo{pages}{968--980}.
\newblock
\showISBNx{978-1-4503-6768-4}
\href{https://doi.org/10.1145/3324884.3416590}{doi:\nolinkurl{10.1145/3324884.3416590}}


\bibitem[White et~al\mbox{.}(2024)]%
        {white2024:livebench}
\bibfield{author}{\bibinfo{person}{Colin White}, \bibinfo{person}{Samuel
  Dooley}, \bibinfo{person}{Manley Roberts}, \bibinfo{person}{Arka Pal},
  \bibinfo{person}{Benjamin Feuer}, \bibinfo{person}{Siddhartha Jain},
  \bibinfo{person}{Ravid {Shwartz-Ziv}}, \bibinfo{person}{Neel Jain},
  \bibinfo{person}{Khalid Saifullah}, \bibinfo{person}{Sreemanti Dey},
  \bibinfo{person}{{Shubh-Agrawal}}, \bibinfo{person}{Sandeep~Singh Sandha},
  \bibinfo{person}{Siddartha~Venkat Naidu}, \bibinfo{person}{Chinmay Hegde},
  \bibinfo{person}{Yann LeCun}, \bibinfo{person}{Tom Goldstein},
  \bibinfo{person}{Willie Neiswanger}, {and} \bibinfo{person}{Micah Goldblum}.}
  \bibinfo{year}{2024}\natexlab{}.
\newblock \showarticletitle{{{LiveBench}}: A Challenging, Contamination-Limited
  {{LLM}} Benchmark}. In \bibinfo{booktitle}{\emph{The {{Thirteenth
  International Conference}} on {{Learning Representations}}}}.
\newblock
\urldef\tempurl%
\url{https://openreview.net/forum?id=sKYHBTAxVa}
\showURL{%
\tempurl}


\bibitem[Xia et~al\mbox{.}(2023)]%
        {xia2023:automated}
\bibfield{author}{\bibinfo{person}{Chunqiu~Steven Xia},
  \bibinfo{person}{Yuxiang Wei}, {and} \bibinfo{person}{Lingming Zhang}.}
  \bibinfo{year}{2023}\natexlab{}.
\newblock \showarticletitle{Automated Program Repair in the Era of Large
  Pre-Trained Language Models}. In \bibinfo{booktitle}{\emph{Proceedings of the
  45th {{International Conference}} on {{Software Engineering}}}}
  \emph{(\bibinfo{series}{{{ICSE}} '23})}. \bibinfo{publisher}{IEEE Press},
  \bibinfo{address}{Melbourne, Victoria, Australia},
  \bibinfo{pages}{1482--1494}.
\newblock
\showISBNx{978-1-6654-5701-9}
\href{https://doi.org/10.1109/ICSE48619.2023.00129}{doi:\nolinkurl{10.1109/ICSE48619.2023.00129}}


\bibitem[Xia and Zhang(2022)]%
        {xia2022:less}
\bibfield{author}{\bibinfo{person}{Chunqiu~Steven Xia} {and}
  \bibinfo{person}{Lingming Zhang}.} \bibinfo{year}{2022}\natexlab{}.
\newblock \showarticletitle{Less Training, More Repairing Please: Revisiting
  Automated Program Repair via Zero-Shot Learning}. In
  \bibinfo{booktitle}{\emph{30th {{ACM Joint European Software Engineering
  Conference}} and {{Symposium}} on the {{Foundations}} of {{Software
  Engineering}}}}. \bibinfo{publisher}{ACM}, \bibinfo{address}{Singapore
  Singapore}, \bibinfo{pages}{959--971}.
\newblock
\showISBNx{978-1-4503-9413-0}
\href{https://doi.org/10.1145/3540250.3549101}{doi:\nolinkurl{10.1145/3540250.3549101}}


\bibitem[Xiang et~al\mbox{.}(2024)]%
        {xiang2024:how}
\bibfield{author}{\bibinfo{person}{Jiahong Xiang}, \bibinfo{person}{Xiaoyang
  Xu}, \bibinfo{person}{Fanchu Kong}, \bibinfo{person}{Mingyuan Wu},
  \bibinfo{person}{Zizheng Zhang}, \bibinfo{person}{Haotian Zhang}, {and}
  \bibinfo{person}{Yuqun Zhang}.} \bibinfo{year}{2024}\natexlab{}.
\newblock \bibinfo{title}{How Far Can We Go with Practical Function-Level
  Program Repair?}
\newblock
\showeprint[arxiv]{2404.12833}~[cs]
\href{https://doi.org/10.48550/arXiv.2404.12833}{doi:\nolinkurl{10.48550/arXiv.2404.12833}}


\bibitem[Xue et~al\mbox{.}(2024)]%
        {xue2024:multiprogramming}
\bibfield{author}{\bibinfo{person}{Tengfei Xue}, \bibinfo{person}{Xuefeng Li},
  \bibinfo{person}{Tahir Azim}, \bibinfo{person}{Roman Smirnov},
  \bibinfo{person}{Jianhui Yu}, \bibinfo{person}{Arash Sadrieh}, {and}
  \bibinfo{person}{Babak Pahlavan}.} \bibinfo{year}{2024}\natexlab{}.
\newblock \bibinfo{title}{Multi-{{Programming Language Ensemble}} for {{Code
  Generation}} in {{Large Language Model}}}.
\newblock
\showeprint[arxiv]{2409.04114}~[cs]
\href{https://doi.org/10.48550/arXiv.2409.04114}{doi:\nolinkurl{10.48550/arXiv.2409.04114}}


\bibitem[Yang et~al\mbox{.}(2024)]%
        {yang2024:revisiting}
\bibfield{author}{\bibinfo{person}{Aidan Z.~H. Yang}, \bibinfo{person}{Sophia
  Kolak}, \bibinfo{person}{Vincent~J. Hellendoorn}, \bibinfo{person}{Ruben
  Martins}, {and} \bibinfo{person}{Claire~Le Goues}.}
  \bibinfo{year}{2024}\natexlab{}.
\newblock \bibinfo{title}{Revisiting Unnaturalness for Automated Program Repair
  in the Era of Large Language Models}.
\newblock
\showeprint[arxiv]{2404.15236}~[cs]
\href{https://doi.org/10.48550/arXiv.2404.15236}{doi:\nolinkurl{10.48550/arXiv.2404.15236}}


\bibitem[Yuan et~al\mbox{.}(2021)]%
        {yuan2021:bartscore}
\bibfield{author}{\bibinfo{person}{Weizhe Yuan}, \bibinfo{person}{Graham
  Neubig}, {and} \bibinfo{person}{Pengfei Liu}.}
  \bibinfo{year}{2021}\natexlab{}.
\newblock \showarticletitle{{{BARTScore}}: {{Evaluating Generated Text}} as
  {{Text Generation}}}. In \bibinfo{booktitle}{\emph{Advances in {{Neural
  Information Processing Systems}}}}, Vol.~\bibinfo{volume}{34}.
  \bibinfo{publisher}{Curran Associates, Inc.}, \bibinfo{pages}{27263--27277}.
\newblock
\urldef\tempurl%
\url{https://proceedings.neurips.cc/paper_files/paper/2021/hash/e4d2b6e6fdeca3e60e0f1a62fee3d9dd-Abstract.html}
\showURL{%
\tempurl}


\bibitem[Zhang et~al\mbox{.}(2020)]%
        {zhang2020:bertscore}
\bibfield{author}{\bibinfo{person}{Tianyi Zhang}, \bibinfo{person}{Varsha
  Kishore}, \bibinfo{person}{Felix Wu}, \bibinfo{person}{Kilian~Q. Weinberger},
  {and} \bibinfo{person}{Yoav Artzi}.} \bibinfo{year}{2020}\natexlab{}.
\newblock \bibinfo{title}{{{BERTScore}}: {{Evaluating Text Generation}} with
  {{BERT}}}.
\newblock
\showeprint[arxiv]{1904.09675}~[cs]
\href{https://doi.org/10.48550/arXiv.1904.09675}{doi:\nolinkurl{10.48550/arXiv.1904.09675}}


\bibitem[Zhong et~al\mbox{.}(2024)]%
        {zhong2024:practical}
\bibfield{author}{\bibinfo{person}{Wenkang Zhong}, \bibinfo{person}{Chuanyi
  Li}, \bibinfo{person}{Kui Liu}, \bibinfo{person}{Tongtong Xu},
  \bibinfo{person}{Jidong Ge}, \bibinfo{person}{Tegawende~F. Bissyande},
  \bibinfo{person}{Bin Luo}, {and} \bibinfo{person}{Vincent Ng}.}
  \bibinfo{year}{2024}\natexlab{}.
\newblock \showarticletitle{Practical Program Repair via Preference-Based
  Ensemble Strategy}. In \bibinfo{booktitle}{\emph{Proceedings of the
  {{IEEE}}/{{ACM}} 46th {{International Conference}} on {{Software
  Engineering}}}}. \bibinfo{publisher}{ACM}, \bibinfo{address}{Lisbon
  Portugal}, \bibinfo{pages}{1--13}.
\newblock
\showISBNx{979-8-4007-0217-4}
\href{https://doi.org/10.1145/3597503.3623310}{doi:\nolinkurl{10.1145/3597503.3623310}}


\bibitem[Zhou et~al\mbox{.}(2023)]%
        {zhou2023:codebertscore}
\bibfield{author}{\bibinfo{person}{Shuyan Zhou}, \bibinfo{person}{Uri Alon},
  \bibinfo{person}{Sumit Agarwal}, {and} \bibinfo{person}{Graham Neubig}.}
  \bibinfo{year}{2023}\natexlab{}.
\newblock \showarticletitle{{{CodeBERTScore}}: Evaluating Code Generation with
  Pretrained Models of Code}. In \bibinfo{booktitle}{\emph{Proceedings of the
  2023 {{Conference}} on {{Empirical Methods}} in {{Natural Language
  Processing}}}}, \bibfield{editor}{\bibinfo{person}{Houda Bouamor},
  \bibinfo{person}{Juan Pino}, {and} \bibinfo{person}{Kalika Bali}} (Eds.).
  \bibinfo{publisher}{Association for Computational Linguistics},
  \bibinfo{address}{Singapore}, \bibinfo{pages}{13921--13937}.
\newblock
\href{https://doi.org/10.18653/v1/2023.emnlp-main.859}{doi:\nolinkurl{10.18653/v1/2023.emnlp-main.859}}


\end{thebibliography}

\end{document}